\documentclass{article}

\PassOptionsToPackage{numbers, compress}{natbib}

\usepackage[main, final]{neurips_2026}

\usepackage[utf8]{inputenc} 
\usepackage[T1]{fontenc}    
\usepackage{hyperref}       
\usepackage{url}            
\usepackage{booktabs}       
\usepackage{amsfonts}       
\usepackage{nicefrac}       
\usepackage{microtype}      
\usepackage{wrapfig}
\usepackage{graphicx}
\usepackage{subcaption}
\usepackage{longtable}
\usepackage{array}
\usepackage{multirow} 
\usepackage{wrapfig}
\usepackage{makecell}
\usepackage{marvosym}
\usepackage[normalem]{ulem}

\usepackage[table]{xcolor}
\usepackage{tcolorbox}
\tcbuselibrary{breakable}
\newtcolorbox{promptbox}[2][Prompt]{
  colback=black!5!white,
  arc=5pt, 
  boxrule=0.5pt,
  fonttitle=\bfseries,
  title=#1, 
  before upper={
    \small
    \obeylines
  },
  fontupper=\fontfamily{ptm}\selectfont,
  colframe=#2,
  breakable,
  lines before break=1,
}
\definecolor{specifications}{rgb}{0.30, 0.72, 0.80}
\definecolor{sys}{rgb}{0.30, 0.72, 0.80}
\definecolor{selector}{rgb}{0.30, 0.72, 0.80}
\definecolor{llm explore}{rgb}{0.30, 0.72, 0.80}
\usepackage{pifont}
\usepackage{colortbl} 
\usepackage{pgfplots}
\pgfplotsset{compat=1.18}
\definecolor{fillcolor}{HTML}{3b82f6}
\definecolor{strokecolor}{HTML}{1e3a8a}

\usepackage{amsmath}
\usepackage{amssymb}
\usepackage{mathtools}
\usepackage{amsthm}
\usepackage[capitalize,noabbrev]{cleveref}
\usepackage{algorithm}
\usepackage{algorithmic}
\usepackage{wrapfig}
\DeclareMathOperator*{\argmax}{arg\,max} 
\usepackage{bm}

\usepackage[T1]{fontenc} 

\title{WebNavigator: Global Web Navigation via Interaction Graph Retrieval}

\author{%
  \textbf{Xuanwang Zhang}\textsuperscript{1,2,3,*} \quad
  \textbf{Yuteng Han}\textsuperscript{1,2,*} \quad
  \textbf{Jinnan Qi}\textsuperscript{1,2} \quad
  \textbf{Xinyu Liu}\textsuperscript{1,2}
  \\
  \textbf{Kunze Meng}\textsuperscript{1,2} \quad
  \textbf{Mulong Xie}\textsuperscript{3} \quad
  \textbf{Zhen Wu}\textsuperscript{1,2,$\dagger$} \quad
  \textbf{Xinyu Dai}\textsuperscript{1,2}
  \\[0.5em]
  \textsuperscript{1}National Key Laboratory for Novel Software Technology,
  Nanjing University, China
  \\
  \textsuperscript{2}School of Artificial Intelligence,
  Nanjing University, China
  \\
  \textsuperscript{3}Fellou AI
  \\[0.5em]
  \textsuperscript{*}Equal contribution. \quad
  \textsuperscript{$\dagger$}Corresponding author.
  \\
  \Letter\ 
  \href{mailto:zhangxuanwang@smail.nju.edu.cn}{\texttt{zhangxuanwang@smail.nju.edu.cn}}
  \quad
  \href{mailto:wuz@nju.edu.cn}{\texttt{wuz@nju.edu.cn}}
  \\[0.4em]
  \href{https://fate-ubw.github.io/webNavigator_homepage/}{%
    \raisebox{-0.08em}{\includegraphics[height=0.9em]{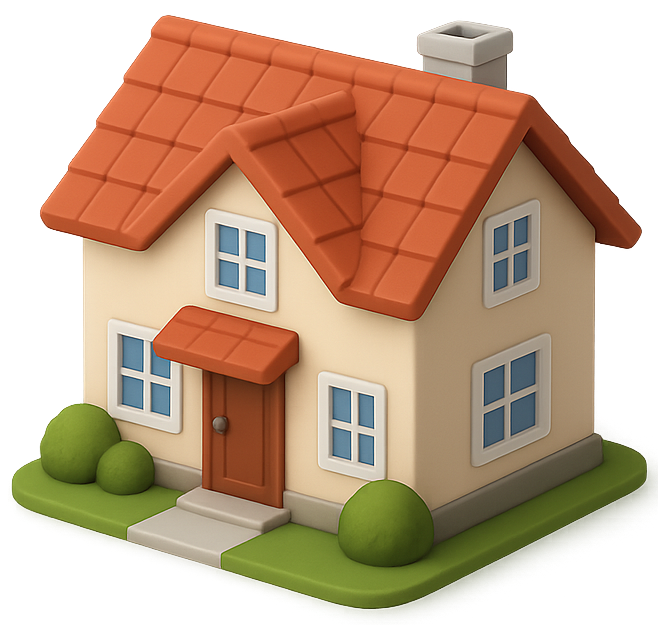}}%
    \enspace\texttt{Project Page}}
  \qquad
  \href{https://github.com/fate-ubw/webNavigator}{%
    \raisebox{-0.15em}{\includegraphics[height=1.05em]{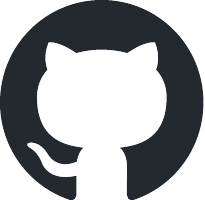}}%
    \enspace\texttt{GitHub}}
}

\begin{document}

\maketitle

\begin{abstract}
  Despite significant advances in autonomous web navigation, current methods remain far from human-level performance in complex web environments. We argue that this limitation stems from Topological Blindness, where agents are forced to explore via trial-and-error without access to the global topological structure of the environment. To overcome this limitation, we introduce WebNavigator, which reframes web navigation from probabilistic exploration into deterministic retrieval and pathfinding. WebNavigator constructs Interaction Graphs via zero-token cost heuristic exploration offline and implements a Retrieve-Reason-Teleport workflow for global navigation online. WebNavigator achieves state-of-the-art performance on WebArena and Online-Mind2Web. On WebArena multi-site tasks, WebNavigator achieves a 72.9\% success rate, more than doubling the performance of enterprise-level agents. This work reveals that Topological Blindness, rather than model reasoning capabilities alone, is an underestimated bottleneck in autonomous web navigation.
\end{abstract}

\section{Introduction}
\label{sec:intro}
Despite achieving superhuman proficiency in automated code generation and complex mathematical theorem proving \cite{openhand, alphaproof}, autonomous agents continue to struggle with human-level performance in dynamic web navigation, particularly in complex cross-site scenarios \cite{webarena2023,webvoyager2024,workarena2024}. Web navigation agents operate by receiving a natural language intent and issuing a sequence of actions, observing only the current page state at each step \cite{webarena2023}. The predominant approach, exemplified by ReAct \cite{react2023}, follows a single linear trajectory where the agent selects each action based solely on the current observation and interaction history, without any awareness of the broader environment. The research community has converged on two dominant paradigms to address this limitation.

\textbf{Paradigm 1: Online Exploratory Search.} Methods in this paradigm attempt to expand an agent's observational scope at inference time through best-first search, Monte Carlo Tree Search, or reflection-guided planning \cite{treesearch2025, exact2025, branch-and-browse2025}. While these approaches acquire broader environmental context than reactive baselines, they remain constrained to local subtree expansion. Each expansion step extends the agent's planning horizon incrementally by accumulating observations through interaction history, without ever revealing the global topological structure of the environment. 
As a result, these methods represent sophisticated trial-and-error within a bounded horizon rather than principled global planning.

\textbf{Paradigm 2: Learned Internal Planning and World Models.} 
A second line of work attempts to internalize environmental knowledge through model parameters or learned world models. Methods in this category train transition models to predict state changes~\cite{webdreamer2025} or elicit latent environmental knowledge from LLM priors~\cite{step2023,agentoccam2025,plan-and-act2025}. 
Recent studies reveal that learned world models are prone to counterfactual hallucinations, fabricating page elements and states that do not exist, comprising the largest share of prediction failures, while methods relying on parametric priors find that leading LLMs perform near random chance when predicting web interaction outcomes from internal knowledge alone~\cite{worldmodelagent2025, finegrainedplanning2025}.
We argue that this brittleness stems from a fundamental information deficit. 
Because the global navigational structure of a website remains latent behind individual page observations, models are forced to fill this gap through parametric approximation rather than grounded environmental knowledge.

Despite these efforts, both paradigms continue to fall short of human-level performance on complex web navigation tasks. The academic community often attributes such shortcomings to insufficient multi-step reasoning capabilities.
However, we argue this view is incomplete. Beyond reasoning capacity, both paradigms share a more fundamental and systematically overlooked structural bottleneck, which we term Topological Blindness. 
Specifically, agents are forced to plan with only a local information subset, comprising current observations, interaction history, and internal priors, while remaining blind to the global topological structure of the website. 

\begin{figure*}[!ht]
  \centering
    \centerline{\includegraphics[width=1.0\textwidth]{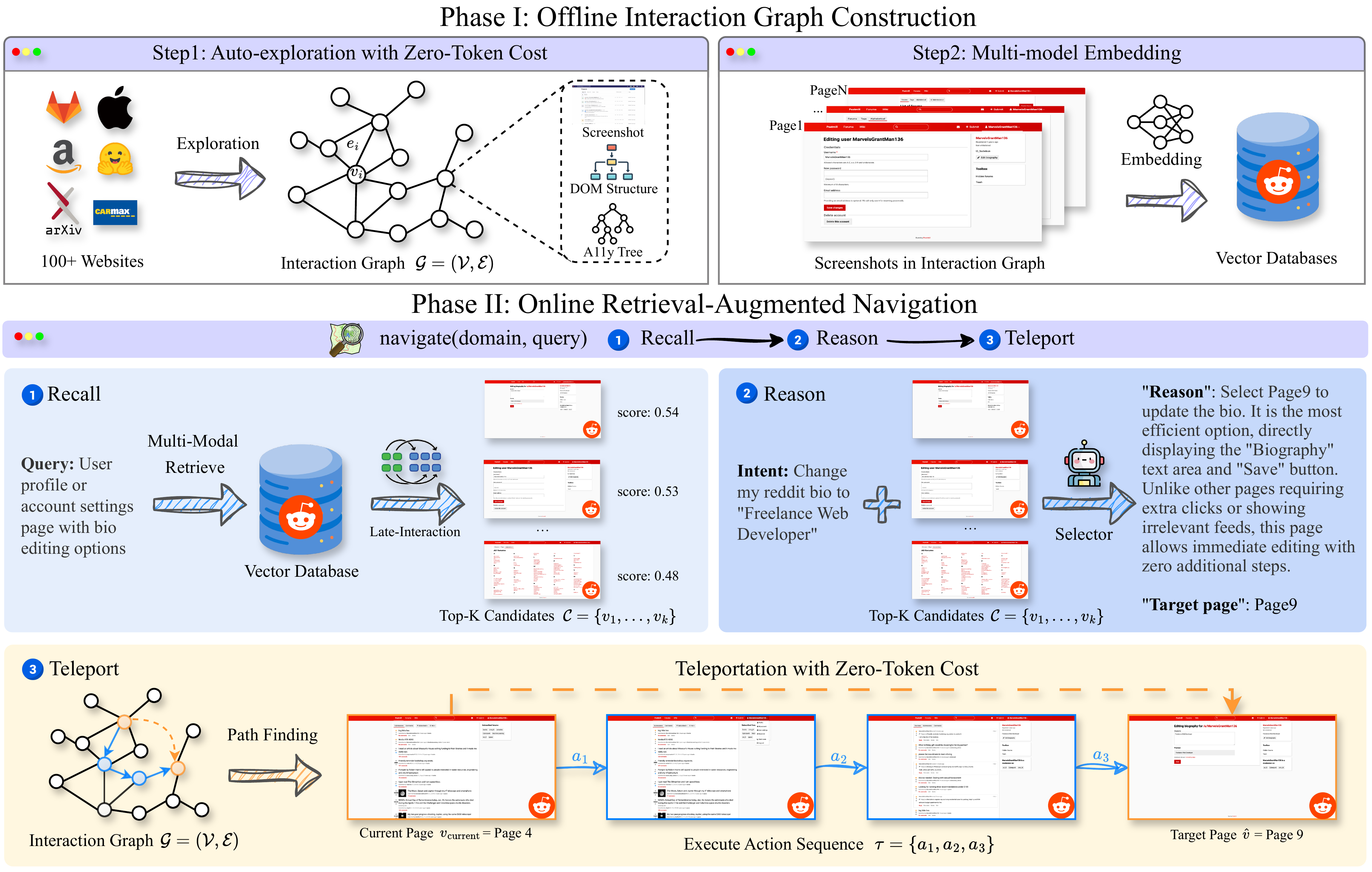}}
    \caption{
Overview of WebNavigator. WebNavigator resolves Topological Blindness via a two-phase paradigm. (1) \textbf{Offline Interaction Graph Construction}. 
A heuristic auto-exploration engine discovers dynamic page observations at zero-token cost and indexes all observations into a vector database. (2) \textbf{Online Retrieval-Augmented Navigation}. The Global-View Navigator implements a three-stage workflow: \textbf{Retrieve} top-$k$ candidates from the Interaction Graph via multimodal retrieval; \textbf{Reason} to identify the optimal target page; and \textbf{Teleport} by computing and executing the shortest path within the Interaction Graph, achieving globally optimal navigation.      
    }
    \label{fig:webnavigator}
\end{figure*}

We posit that web agents, much like expert human users, require a persistent mental map of the environment to achieve optimal planning. 
Even domain experts cannot efficiently navigate an unfamiliar enterprise system on their first encounter. 
Once a human expert has pre-explored a system, navigation becomes deterministic. The expert knows precisely which pages exist, how to reach them, and which path leads most directly to the goal. This observation motivates our central insight. In structured web environments, navigation should be reframed from a probabilistic reasoning challenge into a deterministic retrieval and pathfinding problem.
We introduce WebNavigator, a two-phase framework comprising Offline Interaction Graph Construction and Online Retrieval-Augmented Navigation.
\textbf{Offline Interaction Graph Construction:}
WebNavigator employs a heuristic engine to systematically formalize the website's topological structure into a directed Interaction Graph $\mathcal{G}$. Traditional crawlers parse only static hyperlinks and cannot discover page states that emerge through user interaction, whereas our engine actively interacts with dynamic elements to capture the full navigable structure of the website.
Crucially, this engine builds the Interaction Graph with zero-token cost, requiring no LLM involvement and only a homepage URL as input. Subsequently, all nodes are embedded and indexed into a vector database \cite{jinaembeddingv4}, transforming the Interaction Graph into a retrievable knowledge base.
\textbf{Online Retrieval-Augmented Navigation:}
During inference, the Global-View Navigator (\cref{fig:webnavigator}) implements a three-stage workflow that locates the optimal target page and teleports the agent to it:
(1) \textbf{Retrieve.} The agent issues a navigation query, and the Navigator retrieves the top-$k$ relevant observations from the pre-constructed knowledge base.
(2) \textbf{Reason.} A multimodal selector identifies the optimal observation. 
(3) \textbf{Teleport.} A pathfinding algorithm computes and executes the shortest trajectory to the target observation at zero-token cost.
WebNavigator shifts agents from blind exploration to global planning with only six actions. The Global-View Navigator serves as a modular component that can be integrated into various existing frameworks.

We evaluate WebNavigator on WebArena and Online-Mind2Web \cite{webarena2023,onlinemind2web}. WebNavigator achieves state-of-the-art performance in both WebArena and Online-Mind2Web. Our results demonstrate that WebNavigator achieves a 50.0\% success rate on the most challenging multi-site tasks in WebArena, doubling the previous SOTA under identical configurations. Remarkably, leveraging Gemini-2.5-Pro, WebNavigator establishes a new performance ceiling of 72.9\% on multi-site tasks, more than doubling the enterprise-level agent CUGA~\cite{cuga}. Furthermore, our experiments across 136 real-world websites in Online-Mind2Web confirm the robust generalization of our paradigm. Collectively, these results identify Topological Blindness as a fundamental and overlooked bottleneck that constrains the potential of autonomous web agents.

\section{ Problem formulation}

We formalize the web navigation task as a Partially Observable Markov Decision Process (POMDP), defined by the tuple $\langle \mathcal{S}, \mathcal{A}, \mathcal{O}, \mathcal{T}, \mathcal{R} \rangle$ \cite{webarena2023, exact2025}, where $\mathcal{S}$ denotes the state space, $\mathcal{A}$ the action space, $\mathcal{O}$ the observation space, $\mathcal{T}$ the transition function, and $\mathcal{R}$ the reward function.
An agent receives an intent $i$ and an initial observation $o_0$. At each time step $t$, the agent generates an action $a_t \in \mathcal{A}$ based on the current state $s_t$, where $s_t$ typically encodes the intent $i$, current observation $o_t \in \mathcal{O}$, and interaction history $h_t = (o_0, a_0, \ldots, a_{t-1})$:
\begin{equation}
a_t \sim \pi_{\theta}(a_t \mid i, o_t, h_t)
\end{equation}
where $\theta$ represents the internal knowledge encoded in LLM parameters. After executing action $a_t$, the environment transitions to a new underlying state $s_{t+1} \in \mathcal{S}$ according to the transition function $\mathcal{T}(s_t, a_t)$ and returns a subsequent observation $o_{t+1}$ to the agent. This process continues until the agent issues a termination command or exceeds the step limit, receiving a reward $r \in \{0, 1\}$ based on the functional correctness of task completion \cite{webarena2023}.

We define \textbf{Topological Blindness} as the condition where $\pi_{\theta}$ relies solely on the local information subset $\{o_t, h_t, \theta \}$ without access to the complete observation space $\mathcal{O}$ and transition function $\mathcal{T}$. This structural blindness forces agents into inefficient trial-and-error exploration. We argue that augmenting the policy with environmental knowledge resolves this limitation:
\begin{equation}
a_t \sim \pi_{\theta}(a_t \mid i, o_t, h_t, \mathcal{O}, \mathcal{T})
\end{equation}
With complete environmental information, agents can in principle achieve globally optimal planning within model capacity bounds.
However, providing complete $\mathcal{O}$ and $\mathcal{T}$ to agents remains impractical in real-world scenarios. This raises our central question: \textit{Can we construct a compact environmental representation that approximates $(\mathcal{O}, \mathcal{T})$ to enable global planning in web navigation?}

\section{WebNavigator}
\label{sec:method}
We decompose web navigation into the exploration and local execution stages \cite{plan-and-act2025}. The exploration stage involves navigating across web pages to locate task-relevant observations, such as exploring from the homepage to find the product editing page. The local execution stage involves performing specific interactions within identified pages, such as filling out forms to update product information. Complex tasks alternate between these two stages multiple times. In this paper, we focus on achieving globally optimal planning in the exploration phase.

Previous work has shown that modeling observation transitions $T': (o_t, a_t) \rightarrow o_{t+1}$ is sufficient to simulate web environments \cite{worldmodelagent2025}.
Building on this insight, we introduce the \textbf{Interaction Graph} $\mathcal{G} = (\mathcal{V}, \mathcal{E})$ to capture observation transitions. Nodes $v \in \mathcal{V}$ represent unique page observations, directed edges $e = (v, a, v') \in \mathcal{E}$ represent observation transitions, where executing action $a \in \mathcal{A}$ at node $v$ induces a transition to node $v'$. 
Each node $v$ is grounded via an observation mapping $\phi$, where $o_v = \phi(v) = (o_v^{\text{vis}}, o_v^{\text{str}})$ denotes the multimodal observation comprising a screenshot $o_v^{\text{vis}}$ and structural metadata $o_v^{\text{str}}$ including DOM trees and accessibility trees.
Given the Interaction Graph $\mathcal{G}$ that encodes the observation space $\mathcal{O}$ and transition dynamics $T'$, agents can achieve globally navigation in the exploration phase:
\begin{equation}
a_t \sim \pi_{\theta}(a_t \mid i, o_t, h_t, \mathcal{G})
\end{equation}
By conditioning on $\mathcal{G}$, agents convert web navigation from probabilistic exploration into deterministic pathfinding, effectively addressing Topological Blindness. WebNavigator instantiates this paradigm through two phases: Offline Interaction Graph Construction and Online Retrieval-Augmented Navigation.

\subsection{Phase I:Offline Interaction Graph Construction}
\label{sec:phase1}
Constructing Interaction Graphs $\mathcal{G}$ requires discovering dynamic web states that emerge through user interactions. 
Traditional crawlers parse static hyperlinks to discover URL-addressable pages but miss interaction-triggered states, thereby capturing only a narrow subset of the complete observation space $\mathcal{O}$ \cite{webwalker2025}. Recent LLM-based exploration methods \cite{webatlas2025} suffer from probabilistic reasoning and context-window constraints, leading to redundant revisiting, incomplete coverage, and prohibitive token costs.

\textbf{Heuristic Auto-Exploration Engine}. As shown in \cref{fig:webnavigator} phase I, we develop a heuristic auto-exploration engine based on breadth-first search (BFS) that systematically interacts with dynamic elements to explore the observation space $\mathcal{O}$.
Each discovered node $v \in \mathcal{V}$ is uniquely indexed by hashing its structural components $\text{Hash}(o^{\text{str}}_v, \text{url}_v)$, where each web element is identified via deterministic XPath rather than dynamically assigned BIDs, guaranteeing stable interaction across browser sessions.\footnote{XPath is a W3C standard language: \url{https://www.w3.org/TR/xpath/}} The engine captures comprehensive representations including DOM structures, accessibility trees, and screenshots.
Naive BFS is inefficient because web interactions typically induce only local changes. Child pages inherit most elements from their parents, resulting in redundant exploration. Inspired by the human strategy of avoiding re-clicking explored elements, we design an Adaptive BFS algorithm that leverages structural differencing between DOM trees. At each exploration step, we compute the structural difference between the current node $v$ and its parent $v_{\text{parent}}$ to identify newly added elements. We then extract interactive elements from these differential elements. This strategy significantly reduces exploration overhead by focusing on newly emerged interactive elements, which are substantially fewer than the total number of elements in practice. 
To reach interaction-triggered nodes such as toggled menus, the algorithm simultaneously constructs $\mathcal{G}$ while utilizing it for pathfinding. 
Additionally, the engine employs a configurable block list to exclude hazardous operations and external links from exploration.
Complete details are provided in Appendix~\cref{appendix:algorithm}.

\textbf{Graph Indexing for Retrieval.}
Upon completing exploration, all nodes $v \in \mathcal{V}$ are embedded and indexed in a persistent database for multimodal retrieval during online navigation. Details about retrieval and navigation integration are described in \cref{sec:phase2}.

\subsection{Phase II: Online Retrieval-Augmented Navigation}
\label{sec:phase2}
Directly providing the complete Interaction Graph $\mathcal{G}$ to LLMs is impractical due to context window constraints. Summarizing observations via LLMs not only adds token overhead but also risks losing critical information.
WebNavigator resolves this through the Global-View Navigator, which implements a three-stage workflow: (1) \textbf{Retrieve:} identifying top-$k$ candidate nodes $\mathcal{C} \subset \mathcal{V}$ via multimodal retrieval given navigation query $q$, (2) \textbf{Reason:} selecting the optimal target node $v^* \in \mathcal{C}$ from the candidates, and (3) \textbf{Teleport:} computing and executing the shortest path to $v^*$.

\textbf{Retrieve.} Recent studies in optical compression have demonstrated that vision-language models can process rendered text with significantly fewer tokens than text-only models without sacrificing reasoning performance \cite{glyph2025, deepseekocr}. Building on this insight, WebNavigator adopts screenshots as optically compressed observations for multimodal retrieval, as they provide more compact representations than prohibitively long and noisy DOM trees.
Given a navigation intent $i$, the agent aims to locate the optimal target observation $o^* \in \mathcal{O}$ where the task can be completed. 
To identify the corresponding node in $\mathcal{G}$, the agent formulates a navigation query $q$ based on intent $i$ to specify the target page functionality (e.g., for intent ``Edit product X's price to \$50'', query ``page to edit product information'') and encodes it into multi-vector representation $\mathbf{q} \in \mathbb{R}^{n \times d}$ using the same multimodal embedding model from Phase I, where $n$ denotes the number of tokens and $d$ is the embedding dimension per token \cite{jinaembeddingv4}.
This query embedding $\mathbf{q}$ is then compared with the pre-indexed screenshot embeddings $\{\mathbf{v}_j \in \mathbb{R}^{m_j \times d}\}_{j=1}^{|\mathcal{V}|}$ by computing the late-interaction similarity score \cite{colbert2020}:
\begin{equation}
s_{\text{late}}(\mathbf{q}, \mathbf{v}_j) = \frac{1}{n} \sum_{i=1}^{n} \max_{\ell \in \{1,\dots,m_j\}} \mathbf{q}_i \cdot \mathbf{v}_{j,\ell}^\top
\end{equation}
where $m_j$ denotes the number of tokens in screenshot, $\mathbf{q}_i \in \mathbb{R}^d$ denotes the $i$-th token embedding in the query, and $\mathbf{v}_{j,\ell} \in \mathbb{R}^d$ denotes the $\ell$-th token embedding in screenshot of node $v_j$.
Unlike dense single-vector retrieval, late-interaction computes fine-grained token-level similarities, enabling more precise semantic matching while maintaining computational efficiency \cite{jinaembeddingv4}.
The retrieval process identifies the top-$k$ candidates $\mathcal{C}$ with the highest similarity scores:
\begin{equation}
\mathcal{C} = \{v_1, \dots, v_k\},\ \text{where} \ v_i = \argmax_{\mathclap{v \in \mathcal{V} \setminus \{v_1, \dots, v_{i-1}\}}} s_{\text{late}}(\mathbf{q}, \mathbf{v})
\end{equation}
\textbf{Reason.} 
Since retrieval is based on semantic similarity, the top-$k$ candidates $\mathcal{C}$ may contain visually similar but functionally distinct pages. To identify the optimal target node, we leverage a multimodal LLM as a zero-shot reasoner (prompt detailed in Appendix~\cref{sec:selector_prompt}) to select the most likely candidate $\hat{v}$ by analyzing the visual observations $\mathcal{O}_{\mathcal{C}} = \{o_{v_j}^{\text{vis}}\}_{j=1}^k$ along with the intent $i$:
\begin{equation}
\hat{v} = \arg\max_{\mathclap{v \in \mathcal{C} \cup \{\emptyset\}}} P_\theta(v \mid i, \mathcal{O}_C)
\end{equation}
where $P_{\theta}$ represents the model's probability over the candidate set $\mathcal{C}$. 
When $\hat{v} = \emptyset$, indicating no candidate sufficiently matches the navigation intent, the Navigator bypasses the Teleport stage and the agent falls back to React-style low-level actions to continue task execution. As shown in \cref{sec:none_returns}, such cases are rare and do not necessarily lead to task failure.
By selecting from the top-$k$ candidates rather than engaging in generative exploration, this approach fundamentally reduces navigation complexity from generation to verification \cite{trainingverify2021,auto-eval2024}.

\textbf{Teleport.} Once the target node $\hat{v}$ is determined, the Global-View Navigator computes the shortest path from the current observation $v_{\text{current}}$ to $\hat{v}$ by invoking the pathfinding mechanism used in \cref{sec:phase1}:
\begin{equation}
\tau = (a_1, \dots, a_m) = \operatorname{ShortestPath}(v_{\text{current}}, \hat{v}, \mathcal{G})
\end{equation}
The Navigator executes this action sequence $\tau$ via XPath-grounded actions, transitioning the agent to the target observation with zero-token cost.

Overall, through the Retrieve-Reason-Teleport workflow, the Global-View Navigator approaches globally optimal planning by transforming probabilistic exploration into deterministic retrieval and pathfinding over the pre-constructed graph.
From the agent's perspective, this entire workflow is abstracted into a single high-level action that accepts a navigation query and enables teleportation to the task-relevant page, providing an LLM-friendly interface that eliminates the need to manage complex navigation logic.

\begin{figure*}[ht]
  \centering
    \centerline{\includegraphics[width=1.0\textwidth]{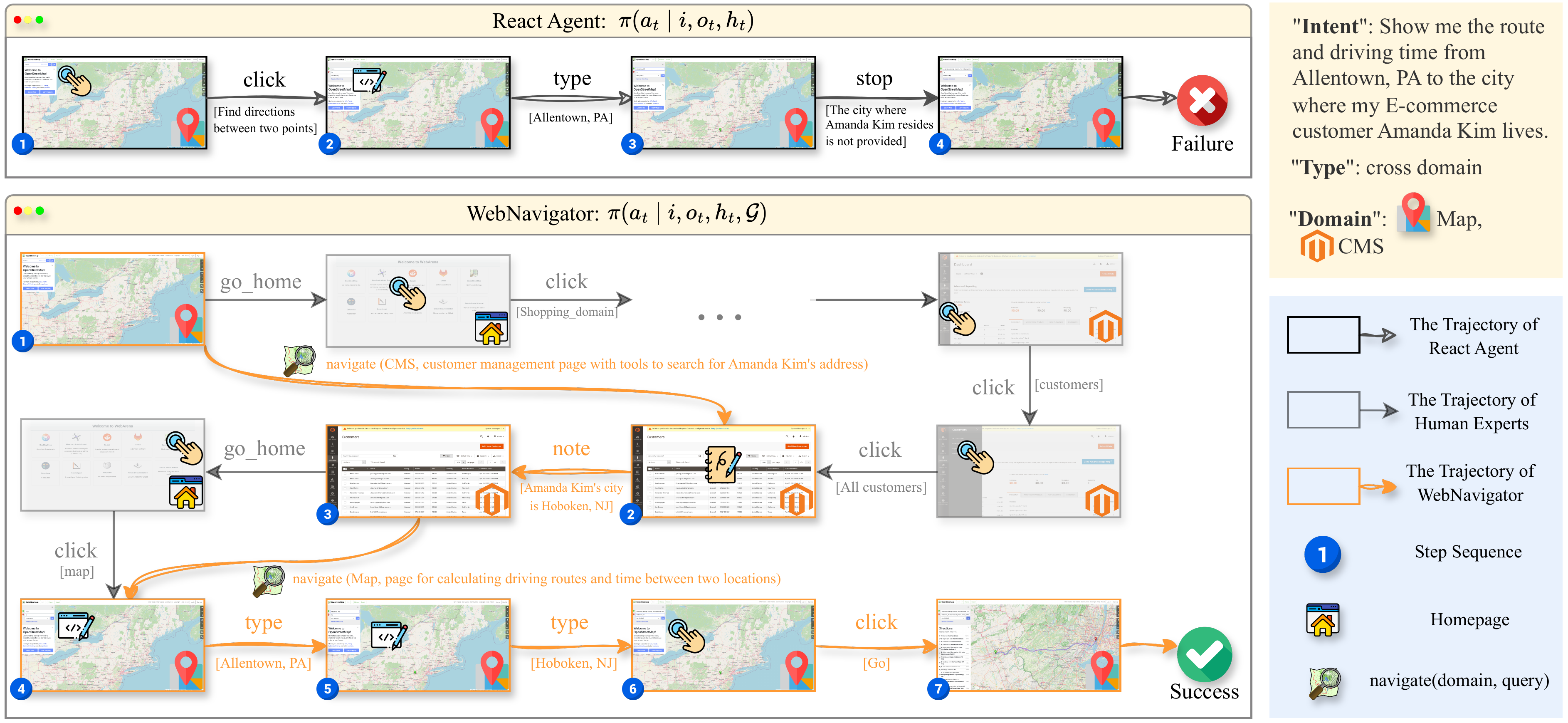}}
    \caption{
      Trajectory comparison on a multi-site task (WebArena 760), which requires retrieving a specific customer address from the CMS to plan a route on the Map. WebNavigator achieves human-level planning via \texttt{navigate(domain, query)} actions, whereas the ReAct baseline prematurely terminates due to Topological Blindness. The human expert trajectory is the gold standard. }
    \label{fig:trajectory}
\end{figure*}

\subsection{WebNavigator Agent Design}
\label{sec:method_agent_design}
Unlike prior approaches that fragment capabilities across numerous atomic primitives \cite{step2023,webarena2023}, we aggregate cross-domain planning, task decomposition, and environment navigation into a high-density yet LLM-friendly interface.
WebNavigator encapsulates the entire Retrieve-Reason-Teleport workflow into a unified action denoted as \texttt{navigate(domain,query)}. This design establishes a paradigm of \textbf{capability aggregation}.
The \texttt{query} parameter enables the agent to focus on one navigation subgoal at a time, naturally decomposing complex tasks into sequential steps. The \texttt{domain} parameter enables cross-domain planning by allowing dynamic switching between Interaction Graphs. The LLM specifies where to navigate, while the underlying workflow handles how to reach the target as a black box.
In contrast to previous methods that fragment tab management capability into explicit actions such as \texttt{tab\_focus}, \texttt{new\_tab}, and \texttt{tab\_close} \cite{webarena2023}, WebNavigator offers a significant advantage by subsuming these capabilities within the Retrieve-Reason-Teleport workflow. The underlying browser state is automatically managed while shielding the LLM from these mechanics.
Methods such as SteP require designing domain-specific actions for each website, including \texttt{find\_commits}  for GitLab and \texttt{search\_customer}  for e-commerce \cite{step2023}.
WebNavigator externalizes environmental knowledge into the Interaction Graph as queryable external memory.
 This design eliminates the need to handcraft specialized actions for every domain, allowing a generic interface to operate uniformly across websites.
Centered on the \texttt{navigate(domain,query)} action, we augment with five primitives for local execution. This yields the most compact action space among existing methods (\cref{table:action_space_comparison}), reducing decision complexity and improving action selection reliability. Complete action definitions and prompts are detailed in Appendix~\cref{sec:action_definition}.

\newcommand{\cellmb}{\cellcolor[HTML]{F5E9DD}}   
\newcommand{\cellml}{\cellcolor[HTML]{D6EAF8}}   

\newcommand{\cellmbb}{\cellcolor[HTML]{FCF4F0}} 
\newcommand{\cellmll}{\cellcolor[HTML]{EBF5FB}} 

\begin{table*}[!t]
\centering
\small
\caption{
Main results on WebArena and Online-Mind2Web. Act \# denotes the action space size. Success rate (SR) for different website domains. \textbf{Bold }and \underline{underlined} values indicate the best and second-best performance among non-enterprise agents. Methods and scores marked with $*$ are our reproduced results. Models marked with $\dagger$ are finetuned.
}
\renewcommand{\arraystretch}{1.2}
\setlength\tabcolsep{3pt}

\resizebox{1\linewidth}{!}{
\begin{tabular}{
    l 
    l 
    c 
    c c c c c c c 
    c             
}

    \toprule
    
    \multirow{4}{*}{\textbf{Method}} & 
    \multirow{4}{*}{\textbf{Model}} & 
    \multirow{4}{*}{\textbf{Act \#}} & 
    \multicolumn{7}{c}{\cellmb \bf WebArena} & 
    \multicolumn{1}{c}{\cellml \bf Online-Mind2Web}\\

    \cmidrule(lr){4-10} \cmidrule(lr){11-11}
    
    ~ & ~ & ~ & 
    \cellmbb \raisebox{0.75em}{\textbf{SR(\%)}} & 
    \cellmbb \raisebox{0.75em}{\textbf{Multisite}} & 
    \shortstack{\includegraphics[height=1.5em]{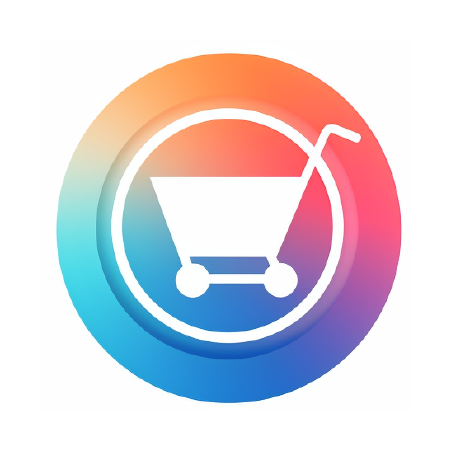} \\ \textbf{Shopping}} & 
    \shortstack{\includegraphics[height=1.5em]{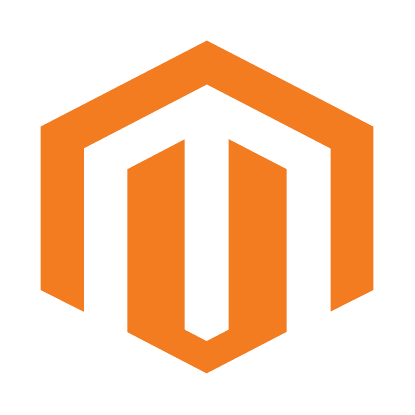} \\ \textbf{CMS}} & 
    \shortstack{\includegraphics[height=1.5em]{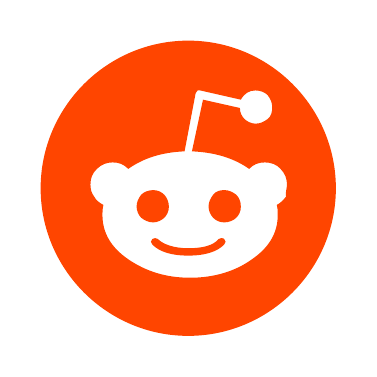} \\ \textbf{Reddit}} & 
    \shortstack{\includegraphics[height=1.5em]{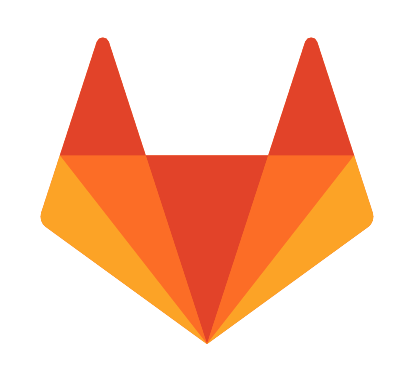} \\ \textbf{GitLab}} & 
    \shortstack{\includegraphics[height=1.5em]{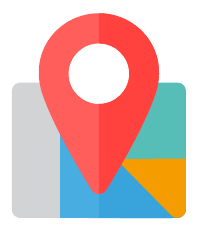} \\ \textbf{Map}} & 
    \cellmll  \raisebox{0.75em}{\textbf{SR(\%)}} \\
    
    \midrule
    Webarena$^*$ \cite{webarena2023}     & GPT-4o & 12 & \cellmbb 21.1 & \cellmbb 8.3 & 24.6 & 20.3 & 22.6 & 22.2 & 18.4 & \cellmll - \\
    Browsergym \cite{browsergym2025}     & GPT-4o & 20 & \cellmbb 31.4 & \cellmbb - &  - & - & - &- &-  & \cellmll - \\
    
    \midrule
    \multicolumn{11}{c}{ \textit{Paradigm 1: Online Exploratory Search} }  \\ \midrule
    
    Tree search \cite{treesearch2025}      & GPT-4o        & 12 & \cellmbb 19.2 & \cellmbb -     & 28.1 & 16.5 & 10.5 & 13.3 & 25.8 & \cellmll 16.6$^*$ \\
    Exact \cite{exact2025}            & GPT-4o        & 16 & \cellmbb -     & \cellmbb -     & -     & -     & -     & 23.5    & -     & \cellmll - \\
    Branch and browse \cite{branch-and-browse2025} & GPT-4o        & 12 & \cellmbb 35.8  & \cellmbb 18.8  & 34.6  & 26.4  & 50.9  & 36.7  & 46.8  & \cellmll 30.3$^*$ \\
    WebPilot  \cite{webploit2025}        & GPT-4o        & -  & \cellmbb 37.2  & \cellmbb -     & 36.9  & 24.7  & 65.1  & 39.4  & 33.9  & \cellmll - \\
    Auto-Eval  \cite{auto-eval2024}       & GPT-4-Preview & 12 & \cellmbb 20.2  & \cellmbb -     & -     & -     & -     & -     & -     & \cellmll 22.0$^*$ \\
    
    \midrule
    \multicolumn{11}{c}{ \textit{Paradigm 2: Learned Internal Planning \& World Models} }   \\ \midrule
     WebDreamer   \cite{webdreamer2025}     & GPT-4o             & 12 & \cellmbb -     & \cellmbb -    & -     & -     & -     & -     & -     & \cellmll 35.0 \\
    WMA  \cite{worldmodelagent2025}   & GPT-4o             & 12 & \cellmbb 16.6  & \cellmbb -    & -     & -     & -     & -     & -     & \cellmll 21.0$^*$ \\
    SteP \cite{step2023}  & GPT-4-Turbo        & 27 & \cellmbb 33.0    & \cellmbb -    & 37.0    & 24.0    & 59.0    & 32.0    & 30.0    & \cellmll 17.4$^*$ \\
    Plan-and-Act  \cite{plan-and-act2025}    & Llama-70B$^{\dagger}$         & 8  & \cellmbb 45.7   & \cellmbb -    & -     & -     & -     & -     & -     & \cellmll - \\
    Agent-e \cite{AgentE2024}    & GPT-4o          & 8  & \cellmbb -  & \cellmbb -    & -     & -     & -     & -     & -     & \cellmll 27.0 \\
    AgentOccam  \cite{agentoccam2025}   & GPT-4-Turbo             & 8  & \cellmbb 43.1 & \cellmbb 14.6 & 40.6 & 45.6 & 61.3 & 37.8 & 46.8 & \cellmll - \\
    AgentOccam$^*$  \cite{agentoccam2025}   & GPT-4o             & 8  & \cellmbb 42.9 & \cellmbb 25.0 & 33.7 & 46.7 & 60.4 & 44.4 & 40.4 & \cellmll 19.6$^*$ \\
    
    \midrule
    \multicolumn{11}{c}{ \textit{Enterprise-Level Autonomous Agents} }   \\ \midrule
    CUGA \cite{cuga} & - & - & \cellmbb 61.7 & \cellmbb 35.4 & 58.3 & 62.6 & 75.5 & 61.7 & 64.2 & \cellmll - \\
    
    \midrule
    \multicolumn{11}{c}{ \textit{New Paradigm: Retrieve and Navigate} }   \\ \midrule
    WebNavigator (Ours)     & Qwen3-VL-32B-Instruct            & 6  & \cellmbb 47.8             & \cellmbb 43.8 & \underline{44.9} & 45.1             & \underline{75.5} & \underline{50.6}& 44.0  & \cellmll 39.7 \\
    WebNavigator (Ours)     & GPT-4o              & 6  & \cellmbb 49.9            & \cellmbb \underline{50.0} & 44.4 & 48.6             & 73.6 & 42.2 & \underline{51.4}  & \cellmll \underline{41.3} \\
    WebNavigator (Ours)     & Claude-Sonnet-4     & 6  & \cellmbb \underline{57.1} & \cellmbb \underline{50.0} & \textbf{51.9}    & \underline{58.2} & \textbf{85.9} & 50.0 & \underline{51.4}  & \cellmll 38.7 \\
    WebNavigator (Ours)     & Gemini-2.5-Pro      & 6  & \cellmbb \textbf{63.3}    & \cellmbb \textbf{72.9}    & \textbf{51.9}             & \textbf{66.5}    & \textbf{85.9} & \textbf{62.2} & \textbf{53.2}  & \cellmll \textbf{52.7} \\
    \bottomrule
    
\end{tabular}
}\label{tab:exp1_results}

\end{table*}

\section{Experiment}

To comprehensively evaluate whether WebNavigator addresses Topological Blindness, we evaluate it on three benchmarks: a controlled environment (WebArena) and two real-world benchmarks (Online-Mind2Web and WebVoyager).

\textbf{Benchmarks.} We primarily evaluate on \textbf{WebArena} \cite{webarena2023}, the most widely used benchmark for web navigation. WebArena comprises 812 carefully designed tasks spanning five representative websites: e-commerce (Shopping), social forums (Reddit), collaborative development (GitLab), content management systems (CMS), and mapping services (Map). WebArena employs programmatic validation mechanisms developed by human experts to assess functional correctness, enabling reliable measurement of task completion. 
To assess generalization capabilities beyond these controlled environments, we extend our evaluation to the real-world benchmark \textbf{Online-Mind2Web} \cite{onlinemind2web}, which consists of 300 diverse tasks across 136 live websites.
We further include \textbf{WebVoyager} \cite{webvoyager2024} as an additional benchmark for end-to-end navigation on live real-world websites.

\textbf{Baselines.} We compare WebNavigator against methods including WebArena \cite{webarena2023}, Browsergym \cite{browsergym2025}, Tree Search \cite{treesearch2025}, Exact \cite{exact2025}, Branch-and-browse \cite{branch-and-browse2025}, WebPilot \cite{webploit2025}, Auto-Eval \cite{auto-eval2024}, WebDreamer \cite{webdreamer2025}, WMA \cite{worldmodelagent2025}, SteP \cite{step2023}, Plan-and-Act \cite{plan-and-act2025}, Agent-e \cite{AgentE2024}, AgentOccam \cite{agentoccam2025} and CUGA \cite{cuga}.

\textbf{Implementation Details}. To ensure a fair comparison, we strictly align our experimental setup with previous works \cite{onlinemind2web,agentoccam2025}. Specifically, we reproduce key baselines using GPT-4o as the unified backbone model.
Detailed experimental parameters are provided in Appendix~\cref{sec:exp_setting}.

\textbf{Main Results}. As shown in \cref{tab:exp1_results}, WebNavigator substantially outperforms previous state-of-the-art methods on WebArena and Online-Mind2Web. Specifically, on WebArena, WebNavigator achieves 47.8\% with Qwen3-VL-32B-Instruct, 49.9\% with GPT-4o, 57.1\% with Claude-Sonnet-4, and 63.3\% with Gemini-2.5-Pro, surpassing WebPilot (37.2\% from Paradigm 1) and Plan-and-Act (45.7\% from Paradigm 2). 
Most significantly, WebNavigator achieves breakthrough performance on multi-site tasks, the most challenging cross-domain setting in WebArena. With GPT-4o and Claude-Sonnet-4, WebNavigator achieves a 50.0\% success rate, surpassing AgentOccam by 100\% relative improvement over its 25.0\%. 
Compared with the enterprise-level agent, WebNavigator with Gemini-2.5-Pro achieves a 72.9\%, more than twice the performance of the CUGA system. Overall, WebNavigator establishes a new performance ceiling on WebArena's multi-site tasks. 
On Online-Mind2Web, WebNavigator achieves 52.7\% with Gemini-2.5-Pro and 41.3\% with GPT-4o, establishing state-of-the-art performance on this challenging benchmark. Crucially, these results validate the generalization capability of WebNavigator across 136 diverse real-world websites.
WebNavigator also outperforms the reproduced WebVoyager agent in our additional real-world evaluation reported in Appendix~\cref{sec:webvoyager_results}.

\subsection{Necessity of the Selector}
\label{sec:selector_necessity}
We evaluate the Selector at both task and query levels. At the task level, we remove the Selector and directly navigate to the top-1 retrieved page. At the query level, we separately measure whether the Retriever recalls a valid target and whether the Selector identifies the final navigation target. Detailed metric definitions and query statistics are provided in Appendix~\cref{sec:selector_details}.
As shown in \cref{tab:selector_ablation,tab:component_analysis}, adding the Selector improves multi-site success by 8.3--18.7 percentage points across all three backbones. The Retriever achieves high Recall@30, exceeding 95\% with Claude-Sonnet-4 and Gemini-2.5-Pro, which shows that the relevant page is highly likely to be included among the retrieved candidates. However, its Top-1 Accuracy remains only 22.97--63.83\%, showing that retrieval alone cannot reliably identify the final target. The Selector directly raises target-selection accuracy from 22.97--63.83\% to 75.82--96.72\%, confirming its importance in the Reason stage.

\begin{table*}[t]
\centering
\begin{minipage}[t]{0.405\textwidth}
\centering
\footnotesize
\setlength{\tabcolsep}{2pt}
\caption{Task-level Selector ablation.}
\label{tab:selector_ablation}
\begin{tabular}{lccc}
\toprule
\textbf{Backbone} & \textbf{Top-1} & \textbf{+ Selector} & \textbf{Gain} \\
\midrule
GPT-4o          & 37.5\% & 50.0\% & +12.5 \\
Claude-Sonnet-4 & 41.7\% & 50.0\% & +8.3 \\
Gemini-2.5-Pro  & 54.2\% & 72.9\% & +18.7 \\
\bottomrule
\end{tabular}
\end{minipage}
\hfill
\begin{minipage}[t]{0.57\textwidth}
\centering
\footnotesize
\setlength{\tabcolsep}{3pt}
\caption{Query-level Selector analysis.}
\label{tab:component_analysis}
\begin{tabular}{lccc}
\toprule
\textbf{Backbone} & \textbf{Top-1 Acc.} & \textbf{Recall@30} & \textbf{Selector Acc.} \\
\midrule
GPT-4o          & 22.97\% & 71.62\% & 75.82\% \\
Claude-Sonnet-4 & 51.02\% & 95.92\% & 96.72\% \\
Gemini-2.5-Pro  & 63.83\% & 95.74\% & 92.00\% \\
\bottomrule
\end{tabular}
\end{minipage}
\end{table*}

\textbf{Structural Foundations of Topological Blindness}. The magnitude of performance improvement across domains directly correlates with the severity of Topological Blindness in each environment. 
Multi-site navigation represents the peak of Topological Blindness, where agents are blind to information in external domains. Consider WebArena task 760: ``Show the route from Allentown, PA to where customer Amanda Kim lives.''
As shown in \cref{fig:trajectory}, without global awareness, the react agent is misled by local cues, leading to premature termination.
In contrast, WebNavigator achieves human-level planning by acquiring cross-domain knowledge, encoded in $\mathcal{G}_{\text{CMS}}$ and $\mathcal{G}_{\text{Map}}$. 
Specifically, WebNavigator teleports to the customer page in CMS to retrieve Amanda Kim's city, then switches to Map to query the driving directions, thus completing the task.

Reddit represents a wide and shallow topology with over 90 forums. React agents can only observe a fraction of paths from the homepage, leading to low-probability guesses. Consider WebArena task 681, which requires posting a GAN-related repository to a relevant subreddit. In this case, an agent lacking global awareness is predisposed to converge on a locally plausible candidate such as \texttt{/f/coolgithubprojects}. In contrast, WebNavigator identifies the optimal targets such as \texttt{/f/deeplearning} and \texttt{/f/MachineLearning} by leveraging the complete forums encoded in $\mathcal{G}_{\text{Reddit}}$.
This global visibility enables WebNavigator to achieve its most significant gain on Reddit.
In deep, complex environments such as CMS, Shopping, and GitLab, agents become trapped by misleading surface pages that mask deeper, task-relevant pages. WebNavigator enables direct navigation to optimal pages, bypassing these traps with substantial gains on Shopping, CMS, and GitLab.
Performance on the Map domain converges across models, where Gemini-2.5-Pro exhibiting no significant margin over GPT-4o or Claude-Sonnet-4. This uniformity is attributed to the constrained observation space of the Map domain ($|\mathcal{V}_{\text{Map}}| = 16$), which mitigates the impact of Topological Blindness. In contrast, other domains contain more than 100 nodes, detailed in \cref{tab:graph_stats}.

\begin{figure*}[t]
  \begin{tikzpicture}[scale=0.85]
    \begin{axis}[
        ybar,
        bar width=10pt,
        width=\textwidth,
        height=4cm,
        scale only axis,
        ymin=60, ymax=80,
        ytick distance=5,
        ylabel={\footnotesize SR (\%)},
        tick align=inside, 
        x tick label style={anchor=north, yshift=-2pt, font=\footnotesize, align=center}, 
        y tick label style={anchor=east, xshift=-2pt, font=\footnotesize},                
        axis lines=box,
        grid=major,
        grid style={gray!30},
        enlarge x limits=0.131,            
        xtick={0.128,1.128,2.128,3.128, 5.413,6.613,7.813, 10.092,11.092,12.092,13.092, 15.275,16.225,17.175},
        xticklabels={1,2,3,4, \text{GPT-4o\qquad}, Gemini-2.5, \text{\qquad Qwen3}, 10,20,30,40, \text{emb-v4\qquad}, emb-v4*, \text{\qquad clip-v2}},
        xticklabel style={font=\footnotesize, align=center},
        nodes near coords={\pgfmathprintnumber[fixed, precision=1, fixed zerofill]{\pgfplotspointmeta}},
        nodes near coords style={
            font=\footnotesize,
            text=black,
            yshift=4pt,
        },
    ]
      \addplot[top color=fillcolor!70, bottom color=fillcolor!30, line width=0.14pt] coordinates {
          (1,63.2) (2,70.8) (3,73.6) (4,75.5)
      };
      \addplot[top color=fillcolor!70, bottom color=fillcolor!30, line width=0.14pt] coordinates {
          (5.7,72.6) (6.9,73.6) (8.1,75.5)
      };
      \addplot[top color=fillcolor!70, bottom color=fillcolor!30, line width=0.14pt] coordinates {
          (9.8,71.7) (10.8,73.6) (11.8,73.6) (12.8,75.5)
      };
      \addplot[top color=fillcolor!70, bottom color=fillcolor!30, line width=0.14pt] coordinates {
          (14.4,73.6) (15.35,67.0) (16.3,66.0)
      };

      \draw[black!30, dashed, thick] (axis cs:4.28, 60) -- (axis cs:4.2, 85);
      \draw[black!30, dashed, thick] (axis cs:9, 60) -- (axis cs:9, 85);
      \draw[black!30, dashed, thick] (axis cs:14.1, 60) -- (axis cs:14.1, 85);

      \node[font=\footnotesize\bfseries, anchor=north] at (axis cs:1.65, 79.8) {(a) Depth};
      \node[font=\footnotesize\bfseries, anchor=north] at (axis cs:6.6, 79.8) {(b) Selector};
      \node[font=\footnotesize\bfseries, anchor=north] at (axis cs:11.61, 79.8) {(c) Top-$k$};
      \node[font=\footnotesize\bfseries, anchor=north] at (axis cs:16.22, 79.8) {(d) Retriever};
    \end{axis}
  \end{tikzpicture}
  \caption{Ablation study on depth, top-$k$, selector model, and retriever on the Reddit domain in WebArena. 
    Unless otherwise specified, the default configuration uses depth=3, top-$k$=30, Gemini-2.5-Flash as the selector, and Jina-embedding-v4 as the retriever.  The Selector group compares GPT-4o, Gemini-2.5-Flash, and Qwen3-VL-8B. The Retriever group compares Jina-embedding-v4, Jina-embedding-v4 (dense), and Jina-clip-v2. $*$ denotes dense embedding mode.} 
  \label{fig:exp3}
\end{figure*}

\section{Analysis}
\subsection{Impact of Environmental Knowledge Completeness and Information Bandwidth}
\label{sec:impact_of_knowledge_completeness}
We investigate the underlying mechanisms that enable WebNavigator to resolve Topological Blindness. Specifically, we conduct controlled experiments on the Reddit domain in WebArena. 

\textbf{Knowledge Completeness}. 
We investigate how the completeness of environmental knowledge affects agent performance. We control knowledge completeness by varying exploration depth from 1 to 4, with fixed GPT-4o, $k=30$, Gemini-2.5-Flash selector, and Jina-embedding-v4 retriever. As shown in the Depth group of~\cref{fig:exp3}, the success rate rises from 63.2\% at depth 1 to 70.8\% at depth 2. This upward trend persists at greater depths, reaching 73.6\% at depth 3 and 75.5\% at depth 4. At depth 1, the Interaction Graph captures only surface-level entry points, leaving many task-relevant pages undiscovered. When depth reaches 2, the Interaction Graph achieves sufficient coverage of task-relevant subreddits, marking a critical transition. Depths 3 and 4 provide additional gains by handling tasks requiring deeper navigation, though with smaller incremental improvements.
These results confirm that sufficient knowledge completeness is necessary to address Topological Blindness.

\textbf{Information Bandwidth}.
We investigate how the information bandwidth affects agent performance. The retrieval top-$k$ parameter controls the number of candidates transferred from the knowledge base to the selector. Larger $k$ expands the decision space of the Selector but increases computational overhead, while insufficient $k$ risks excluding the optimal target. We vary $k$ from 10 to 40 while keeping all other parameters fixed. As shown in the Top-$k$ group of~\cref{fig:exp3}, performance improves as $k$ increases. At $k=40$, WebNavigator achieves 75.5\%, surpassing our main configuration of $k=30$. More importantly, even with the narrowest bandwidth at $k=10$, WebNavigator achieves 71.7\% success rate, outperforming all  non-enterprise baseline methods in \cref{tab:exp1_results}. This robustness under constrained bandwidth demonstrates that knowledge completeness, rather than the retrieval channel width, is the primary determinant of performance.

\textbf{Task Simplification via Verification Asymmetry.} A fundamental principle holds that verifying a solution is inherently easier than generating one~\cite{auto-eval2024, onlinemind2web}. WebNavigator instantiates this principle in web navigation by transforming trajectory generation into candidate selection. Rather than constructing a navigation path from scratch, the agent selects the optimal page from a pre-retrieved candidate set.
We evaluate Selectors ranging from leading proprietary models to the lightweight Qwen3-VL-8B-Instruct.  As shown in the Selector group of~\cref{fig:exp3}, the 8B model achieves a 75.5\% success rate, performing on par with GPT-4o  and Gemini-2.5-Flash. This consistency across model scales confirms that the WebNavigator successfully offloads navigation complexity from model reasoning to structured knowledge retrieval.

\textbf{Retrieval Granularity.}
We examine whether the granularity of multimodal retrieval affects the performance of WebNavigator. Specifically, we compare late interaction retrieval, which leverages token-level alignment via multi-vector representations, against traditional dense methods. We evaluate Jina-embedding-v4 in both late interaction and dense mode (denoted by $*$ in~\cref{fig:exp3}), and Jina-clip-v2 \cite{jinaclipv2025} in dense mode only. As shown in the Retriever group of~\cref{fig:exp3}, late interaction retrieval achieves a 73.6\% success rate, substantially outperforming dense embedding approaches at 66-67\%. Dense embeddings compress entire screenshots into fixed-size vectors, discarding fine-grained spatial cues such as specific buttons, form fields, or menu structures. In contrast, late interaction preserves local visual semantics through token-level matching, enabling the retriever to identify specific page regions that satisfy the navigation intent. This confirms that web navigation requires fine-grained retrieval rather than global compression.

\subsection{Error Analysis}

We manually inspect all failed trajectories of WebNavigator with Gemini-2.5-Pro across multiple WebArena domains, categorizing each failure into one of four types. \textbf{Retrieval Failure} occurs when the Selector returns \texttt{None} or selects an incorrect target page. \textbf{Execution Failure} occurs when the agent reaches the target page but fails due to LLM limitations in local reasoning, including wrong answers, incorrect output format, missed constraints, or misjudged task completion. \textbf{Step Limit Exhaustion} occurs when the agent exhausts the step budget before completing the task. 
\textbf{Annotation/Environment Error} occurs when ground-truth annotations contain ambiguity or errors, or when the environment behaves unexpectedly.
\begin{figure*}  
  \centering
  \includegraphics[width=0.9\textwidth]{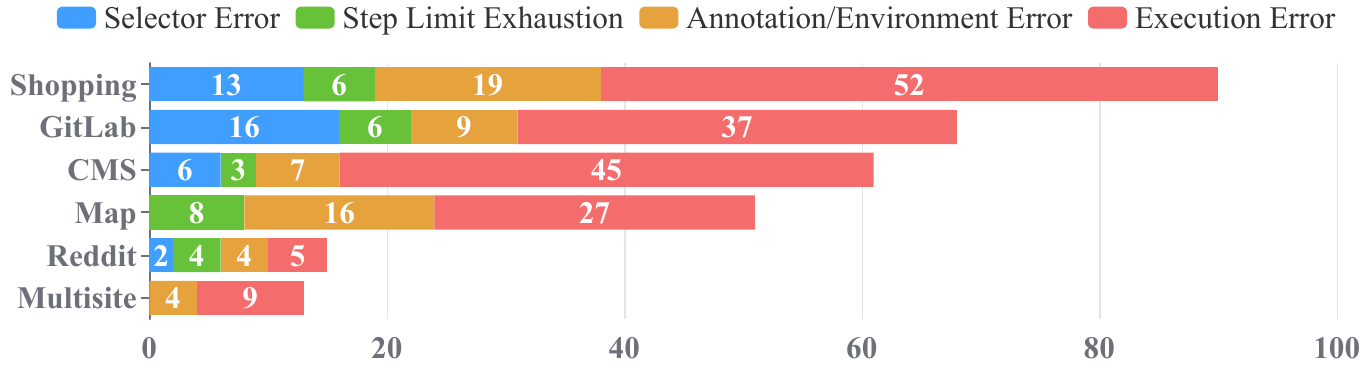}
    \caption{Failure analysis on WebArena (Gemini-2.5-Pro).}
  \label{fig:error_analysis}
\end{figure*}
As shown in \cref{fig:error_analysis}, retrieval failure accounts for only 12.4\% of all failures, indicating that the retrieval and pathfinding pipeline reliably handles navigation across the vast majority of tasks. Notably, multi-site tasks record zero retrieval failures, confirming that cross-domain retrieval is robust even in the most challenging setting. The dominant failure mode is execution failure at 58.7\%, where the agent reaches the correct page but fails due to LLM limitations in local reasoning. Step limit exhaustion accounts for a further 9.1\%, reflecting that even leading models cannot complete local tasks within the budget. Together, LLM-side failures constitute 67.8\% of all failures, confirming that residual failures concentrate on LLM capability limitations rather than framework design. The remaining 19.8\% corresponds to annotation and environment errors, which affect only 7.3\% of all WebArena tasks and fall outside the agent's control.




\section{Related Work}
\label{sec:related_works}
\textbf{Paradigm 1: Online Exploratory Search.} This paradigm attempts to mitigate Topological Blindness through search-based algorithms during task execution. Specifically, these approaches employ best-first search and Monte Carlo Tree Search (MCTS) guided by model-based value functions~\cite{LATS2024,treesearch2025}. To enhance reliability, reflective mechanisms leverage contrastive reflection or multi-agent debate to correct navigation errors~\cite{auto-eval2024, exact2025}. Additionally, other works utilize hierarchical structures to optimize local execution~\cite{webploit2025, branch-and-browse2025}. However, such exploration is transient and environment-agnostic, discarding the acquired structural knowledge after each episode. Unlike these reactive exploration approaches, our framework persists environmental knowledge into a reusable interaction graph.

\textbf{Paradigm 2: Learned Internal Planning and World Models.} This paradigm attempts to mitigate Topological Blindness by learning environmental knowledge within model parameters or building world models. One significant direction is the construction of explicit environmental simulators, which learn world models to predict state transitions~\cite{webdreamer2025, worldmodelagent2025}. Another approach focuses on structural planning, utilizing policy stacks or planning trees to recursively break down long-term objectives~\cite{step2023, agentoccam2025}. To improve quality, specialized architectures separate high-level planning from low-level execution~\cite{plan-and-act2025}. Despite their efficiency, these methods rely on probabilistic inference often yielding hallucinated transitions in novel environments. In contrast, our Retrieve-and-Navigate paradigm bypasses the need for unreliable internal simulations by grounding agent decisions in pre-constructed environmental knowledge.

\section{Conclusion}
This work introduces WebNavigator, a novel paradigm that overcomes Topological Blindness by transforming probabilistic exploration into deterministic retrieval and pathfinding, achieving global planning with only 6 actions. WebNavigator achieves state-of-the-art performance on WebArena and Online-Mind2Web. On WebArena multi-site tasks, WebNavigator demonstrates a performance ceiling of 72.9\%. These results reveal that Topological Blindness represents an underestimated bottleneck in autonomous web navigation.


\clearpage
\bibliographystyle{unsrtnat}   
\bibliography{example_paper}


\newpage
\appendix




\section{Limitations and Societal Impacts}
\subsection{Limitations}
\label{sec:limitations}

\paragraph{Extension to Computer Use and Mobile Environments.}
This work focuses on browser-based web navigation and does not extend to computer use or mobile navigation scenarios. Computer use and mobile 
navigation are pursued as separate research directions within the community, as they involve fundamentally different interaction paradigms, including desktop GUI manipulation and touch-based interfaces. Adapting WebNavigator to these environments is a promising direction for future work.

\paragraph{Browser Harness Coverage of Corner Cases.}
The current browser harness does not yet handle all corner cases that arise in real-world web navigation. For example, observations may be captured before a page has fully loaded, resulting in incomplete accessibility trees. In addition, dynamically injected Browser element IDs (BIDs) may become invalid by the time an action is executed, causing action execution to fail when the target element can no longer be located by its BID. 
Although such cases are rare in practice, improving harness coverage to handle them remains an important direction for future engineering work.

\paragraph{Upfront Graph Construction.}
WebNavigator requires an Interaction Graph to be constructed before task execution on a new website. Although graph construction involves no LLM calls and the resulting graph is fully reusable across tasks, this prerequisite step adds an initial time investment before deployment. 
We plan to address this by developing a dedicated SDK for web developers that automates graph construction and maintenance, making WebNavigator a lightweight infrastructure layer for large-scale web navigation deployment.

\paragraph{Limited Support for HTML canvas elements.} 
WebNavigator relies on accessibility trees for element grounding, which makes it ineffective on websites where core functionality is rendered via HTML canvas elements, such as browser-based games or interactive visualization tools. 
These websites expose no semantic accessibility structure, leaving pure visual approaches as the only viable option. Supporting such environments requires fundamentally different grounding mechanisms and is left for future work.

\paragraph{Manual Selection of Exploration Depth.}
The exploration depth parameter controls the coverage of the Interaction Graph and is currently set manually.
Although Section~\ref{subsec:graph_stats} provides principled guidance for depth selection, an automatic method that adapts depth to each website would further reduce deployment overhead. We leave the development of such an adaptive depth selection mechanism for future work.

\subsection{Broader Impact}
\label{sec:impact}
WebNavigator advances the capability of autonomous web agents to complete complex, goal-oriented tasks across diverse real-world websites, with the potential to assist users in automating daily web-based work. 
However, we acknowledge the potential risks associated with autonomous web interactions, including the possibility of malicious dual-use such as automating CAPTCHA bypassing or unauthorized data scraping. We emphasize that future deployment must incorporate strict usage policies and human-in-the-loop mechanisms, and we call on the research community to establish shared safety standards for autonomous web automation.

\section{Implementation Details and Hyperparameter Settings}
\label{sec:exp_setting}
We conduct all experiments using an RTX A6000 (48GB) server with 8 GPUs.
\cref{tab:global_hyperparams} and \cref{tab:graph_stats} summarize the experimental hyperparameters and per-domain graph construction statistics, respectively. For WebArena, we conduct a systematic exploration up to depth 3, yielding Interaction Graphs that provide comprehensive coverage of the five websites. For Online-Mind2Web, which encompasses 136 diverse real-world websites, we adopt a conservative exploration depth of 1. This strategy ensures broad coverage across all 136 websites within a limited time frame, capturing the majority of functional entry points for each website. 
Despite this conservative setting, WebNavigator achieves SOTA performance as shown in \cref{tab:exp1_results}, demonstrating that even shallow exploration enables effective generalization from controlled environments to real-world websites.
We use the same depth-1 configuration for WebVoyager.

\subsection{Experimental Configuration}
\label{subsec:exp_config}

The global hyperparameters shared across all experiments are listed in \cref{tab:global_hyperparams}. The selector model, retriever, and inference settings remain fixed across all domains and benchmarks.

\begin{table}[htbp]
\centering
\caption{Global hyperparameters for WebNavigator.}
\label{tab:global_hyperparams}
\begin{tabular}{lc}
\toprule
\textbf{Parameter} & \textbf{Value} \\
\midrule
Observation       & Acc. Tree + Image \\
Selector          & Gemini-2.5-Flash \\
Selector Input    & Intent $i$ + top-$k$ candidate screenshots \\
Retriever         & Jina-embedding-v4 \\
Top-$k$           & 30 \\
Temperature       & 0.5 \\
Top-$p$           & 0.95 \\
Max Steps         & 20 \\
\bottomrule
\end{tabular}
\vspace{4pt}
\end{table}

\subsection{Graph Construction and Statistics}
\label{subsec:graph_stats}

\cref{tab:graph_stats} reports the exploration depth, resulting graph size, branching factor, and construction time for each WebArena domain.
Notably, the Interaction Graphs constructed by our adaptive BFS are highly sparse and tree-like, with branching factors ranging from 1.10 to 1.98. 
This sparsity is consistent with modern web design principles, where navigation hierarchies guide users through sequential functional pathways rather than densely interconnected structures.

\begin{table}[htbp]
\centering
\caption{Per-domain graph construction statistics on WebArena.}
\label{tab:graph_stats}
\begin{tabular}{lccccc}
\toprule
\textbf{Parameter} & \textbf{Map} & \textbf{CMS} & \textbf{Shopping} & \textbf{GitLab} & \textbf{Reddit} \\
\midrule
Exploration Depth      & 1   & 2   & 2        & 2      & 3 \\
Graph Nodes            & 16  & 122 & 570      & 838    & 225 \\
Graph Edges            & 19  & 138 & 627      & 1,659  & 295 \\
Branching Factor       & 1.19 & 1.13 & 1.10    & 1.98   & 1.31 \\
Construction Time (min)& 2.1 & 23.1 & 122.1   & 316.8  & 31.1 \\
\bottomrule
\end{tabular}
\end{table}

\paragraph{Selection of Exploration Depth.}
The exploration depth $D$ is set differently per domain, guided by the principle of Broad, Shallow Information Structures~\cite{webdesign1998}, which demonstrates that modern websites concentrate functional entry points at shallow depths. Websites with functionally simple interfaces and shallow navigation, such as Map, require only $D=1$ to capture primary functionality. In contrast, structurally complex platforms with nested features, such as GitLab and Reddit, benefit from $D=2$ or $3$. Consequently, a modest $D$ suffices to cover the vast majority of functional entry points on the website.

\paragraph{Construction Cost.}
Graph construction incurs only a modest one-time offline cost. The most time-consuming domain, GitLab (depth 2), requires 316.8 minutes of commodity CPU wall-clock time with no GPU or LLM dependency. On AWS c7i.xlarge (USD 0.17/hour), constructing the GitLab Interaction Graph costs approximately \$0.90. This one-time investment is shared across all tasks on the same website, eliminating repeated exploration overhead during inference.

\subsection{Selection of Top-$k$}
\label{subsec:topk}

The top-$k$ parameter controls the number of candidate nodes passed from the vector database to the Selector in the Retrieve-Reason-Teleport workflow. Larger $k$ expands the Selector's decision space but increases computational overhead; insufficient $k$ risks excluding the optimal target page.
We set $k=30$ uniformly across all domains and benchmarks. This value provides an optimal balance between retrieval coverage and inference efficiency, as demonstrated by the top-$k$ ablation in \cref{fig:exp3}. A universal $k$ eliminates the need for per-site tuning and generalizes robustly across websites of different scales. Our Online-Mind2Web experiments confirm this design choice, where $k=30$ achieves state-of-the-art performance across 136 real-world websites without any per-site adjustment.

\section{Algorithm Details}
\label{appendix:algorithm}

\cref{alg:exploration} presents the complete procedure for constructing the Interaction Graph $\mathcal{G}$. \textbf{Structural hashing} provides deterministic node identification by hashing of the DOM trees and the URL. This primitive supports core mechanisms that optimize exploration. The algorithm employs \textbf{adaptive structural differencing} (Lines 11-15) to optimize exploration efficiency by computing set differences between parent and child structural representations, focusing only on newly emerged elements. To handle non-URL-accessible nodes, the algorithm features an \textbf{interleaved navigation} mechanism (Lines 8-9) that leverages $\texttt{ShortestPath}$ on the partially constructed graph $\mathcal{G}$.

\textbf{Structural Hashing}. Each element $e$ in the structural observation $o_v^{\text{str}}$ is assigned a deterministic hash based on its full XPath: $h_e = \texttt{MD5}(e_\text{xpath})$.\footnote{XPath is a W3C standard language: \url{https://www.w3.org/TR/xpath/}. The MD5 Message-Digest Algorithm: \url{https://www.rfc-editor.org/rfc/rfc1321}.}The structural representation of a node is defined as the set of all element hashes: $H_v = \{h_e  \mid e \in \text{Elements}(v)\}$. Node identity combines this structural fingerprint with URL metadata: $\text{id}_v = \texttt{MD5}(H_v \| \text{url}_v)$. This design ensures that nodes with identical DOM structures and URLs are identified as the same node, while any structural mutation (element addition, removal, or relocation) yields a distinct hash.

\textbf{Structural Differencing}. Given parent node $v_{\text{parent}}$ and child node $v$, we compute structural difference via set subtraction: $\Delta H_{v} = H_v \setminus H_{v_{\text{parent}}}$. Since both representations are hash sets, this operation executes in expected $O(|H_v|)$ time, avoiding expensive tree alignment algorithms. $\Delta H_{v}$ represents the hashes of elements newly introduced in the child relative to its parent.


\textbf{Why Interleaved Navigation is Necessary}. Unlike explicit graphs where all nodes are directly accessible, many observations lack direct URL access and are reachable only via specific interaction sequences, such as toggling a menu to reveal hidden content. To explore depth $d+1$, the engine must navigate to each parent node $v_i$ at depth $d$. For nodes lacking direct URL access, navigating to $v_i$ requires replaying the action sequence from the root. This path reconstruction requirement introduces the interleaved exploration-navigation paradigm, where the algorithm simultaneously constructs $\mathcal{G}$ while utilizing it for pathfinding. Specifically, the engine constructs depth $d+1$ by iterating through each parent node $v_i$ at depth $d$. 
To reach each $v_i$, the engine computes the shortest path from the root through $\mathcal{G}$ and executes the corresponding action sequence.
Once $v_i$ is materialized in the browser, the engine discovers its children at depth $d+1$.
\cref{tab:notation} summarizes notation used in \cref{alg:exploration} that extends the main text.

\textbf{Block List}. Given that certain websites contain potentially dangerous operations (e.g., account deletion), we employ a Block List mechanism to bypass such hazardous actions during the exploration phase. The block list can be manually configured with regular expression rules to prohibit the exploration of any elements that match the specified criteria.
Additionally, external links are automatically excluded.


\begin{table}[htbp]
  \centering
  \caption{Additional notation for Algorithm~\ref{alg:exploration}}
  \label{tab:notation}
  \begin{tabular}{cc}
  \toprule
  \textbf{Symbol} & \textbf{Description} \\
  \midrule
  $\mathcal{Q}$ & BFS queue organized by exploration depth \\
  $\mathcal{L}_d$ & Set of nodes at depth $d$ \\
  $\mathcal{R}$ & Set of interested accessibility roles (e.g., \texttt{link}, \texttt{button}) \\
  $\mathcal{I}_v$ & Interactive elements extracted from node $v$ \\
  $H_v$ & Structural hash set $\{h_e \mid e \in \text{Elements}(v)\}$ \\
  $\Delta H_{v}$ & Structural hash difference $H_v \setminus H_{v_{\text{parent}}}$ \\
  $\text{id}_v$ & Unique identifier $\texttt{MD5}(H_v\|\text{url}_v)$ \\
  \bottomrule
  \end{tabular}
  \end{table}

\begin{algorithm}[htbp]
\caption{Adaptive BFS Exploration with Interleaved Navigation}
\label{alg:exploration}
\begin{algorithmic}[1]
\REQUIRE Start URL $\text{url}_0$, Maximum depth $D$, Interested roles $\mathcal{R}$
\ENSURE Interaction Graph $\mathcal{G} = (\mathcal{V}, \mathcal{E})$

\STATE $\mathcal{G} \gets (\emptyset, \emptyset)$
\STATE $v_0 \gets \texttt{Snapshot}(\text{url}_0)$ 
\STATE $\mathcal{V} \gets \{v_0\}$; $\mathcal{Q} \gets [\{v_0\}]$

\FOR{$d = 0$ \textbf{to} $D-1$}
\STATE $\mathcal{L}_d \gets \mathcal{Q}.\text{dequeue}()$; $\mathcal{L}_{d+1} \gets \emptyset$
\FOR{\textbf{each} $v \in \mathcal{L}_d$}
\STATE \COMMENT{\textit{--- Interleaved Navigation ---}}
\STATE $\tau \gets \texttt{ShortestPath}(v_0, v, \mathcal{G})$
\STATE $o_v^{\text{str}} \gets \texttt{ExecutePath}(\tau)$
\STATE \COMMENT{\textit{--- Adaptive Structural Differencing ---}}
\STATE $\mathcal{I}_v \gets \texttt{GetInteractiveElements}(o_v^{\text{str}}, \mathcal{R})$
\IF{$\exists\, v_{\text{parent}}$ s.t. $(v_{\text{parent}}, a, v) \in \mathcal{E}$}
\STATE $\Delta H_{v} \gets H_v \setminus H_{v_{\text{parent}}}$
\STATE $\mathcal{I}_v \gets \{e \in \mathcal{I}_v  \mid h_e \in \Delta H_{v}\}$
\ENDIF
\STATE \COMMENT{\textit{--- Explore New Elements ---}}
\FOR{\textbf{each} $e \in \mathcal{I}_v$}
\STATE $a \gets \texttt{CreateAction}(e)$
\STATE $v' \gets \texttt{ExecuteAndSnapshot}(v, a)$
\STATE $\text{id}_{v'} \gets \texttt{MD5}(H_{v'}\|\text{url}_{v'})$
\IF{$\text{id}_{v'} \notin \{\text{id}_u  \mid u \in \mathcal{V}\}$}
\STATE $\mathcal{V} \gets \mathcal{V} \cup \{v'\}$
\STATE $\mathcal{L}_{d+1} \gets \mathcal{L}_{d+1} \cup \{v'\}$
\ENDIF
\STATE $\mathcal{E} \gets \mathcal{E} \cup \{(v, a, v_{\text{target}}) | v_{\text{target}} \in \mathcal{V} \text{, where } \text{id}_{v_{\text{target}}} = \text{id}_{v'}\}$
\ENDFOR
\ENDFOR
\STATE $\mathcal{Q}.\text{enqueue}(\mathcal{L}_{d+1})$
\ENDFOR
\STATE \textbf{return} $\mathcal{G}$
\end{algorithmic}
\end{algorithm}

\begin{table}[htbp]
  \centering
  \caption{Function descriptions for Algorithm~\ref{alg:exploration}}
  \label{tab:functions}
  \begin{tabular}{cm{8cm}}
  \toprule
  \textbf{Function} & \multicolumn{1}{c}{\textbf{Description}} \\
  \midrule
  \texttt{Snapshot} & Captures a comprehensive observation of a specific page state, including both visual and structural representations, and returns an interaction graph node representing this page. \\
  \midrule
  \texttt{ShortestPath} & Computes the shortest action sequence from the start node $v_0$ to target node $v$ through the partially constructed graph $\mathcal{G}$. If no path from $v_0$ to $v$ exists in the graph, it performs a reverse search from $v$ to find the nearest reachable node that is directly accessible via URL, and then derives the path in the forward direction. Returns an ordered action sequence $\tau = (a_1, \ldots, a_m)$.\\
  \midrule
  \texttt{ExecutePath} & Executes a navigation trajectory $\tau$ in the browser. Navigates to the starting node, then sequentially performs each action to reach the target node. \\
  \midrule
  \texttt{GetInteractiveElements} & Extracts all interactive elements matching the specified accessibility roles $\mathcal{R}$ from the structural observation $o_v^{\text{str}}$. \\
  \midrule
  \texttt{CreateAction} & Constructs an action to trigger the given interactive element $e$. \\
  \midrule
  \texttt{ExecuteAndSnapshot} & Executes action $a$ on the web page represented by $v$ and returns the \texttt{Snapshot} of the resulting page state. \\ 
  \bottomrule
  \end{tabular}
\end{table}

\section{Interaction Graph Maintenance and Incremental Update}
The Interaction Graph $\mathcal{G}$ maintains currency with website evolution through a systematic verification and incremental update mechanism. The Heuristic Auto-Exploration Engine periodically executes a verification process that traverses all nodes $v \in \mathcal{V}$ and edges $e \in \mathcal{E}$ in the Interaction Graph. For each node $v$, the engine revisits its corresponding page and validates whether the node remains accessible. Similarly, for each edge $e = (v, a, v') \in \mathcal{E}$, the engine verifies whether executing action $a$ at node $v$ still produces the expected transition to $v'$. The engine removes invalid nodes and edges from $\mathcal{G}$.  For nodes that remain valid after verification, the engine compares the current DOM tree with the cached version to identify newly added elements. The engine then explores from these new elements, adding newly discovered nodes and edges to $\mathcal{G}$. This incremental mechanism minimizes redundant exploration while enabling rapid graph updates. Through this mechanism, the Interaction Graph accurately reflects the latest website topology without requiring LLM involvement.

\section{Detailed Action Space Comparison Across Methods}
We define the action space as the set of actions exposed to the LLM's decision-making context at each step. Under this definition, actions encapsulated within black-box modules do not contribute to the action space size, as the LLM neither observes nor selects them. This distinction is critical for fairly evaluating decision complexity across methods.
WebNavigator exposes only 6 actions to the LLM: \texttt{navigate}, \texttt{click}, \texttt{type}, \texttt{go\_back}, \texttt{note}, and \texttt{stop}. The \texttt{navigate(domain, query)} action encapsulates the entire Retrieve-Reason-Teleport workflow as a black box. While this workflow internally invokes browser primitives such as \texttt{goto\_url} and \texttt{click} during path execution, these are handled deterministically by the pathfinding mechanism and never enter the LLM's context window. In contrast, prior methods such as SteP expose 27 actions to the LLM, including numerous domain-specific actions such as \texttt{find\_commits} and \texttt{search\_customer} that must be handcrafted for each website. WebNavigator subsumes all such domain-specific functionality within the unified \texttt{navigate} interface, reducing decision complexity while maintaining generality across websites.
We list the action space of WebArena~\cite{webarena2023}, BrowserGym~\cite{browsergym2025}, AgentOccam~\cite{agentoccam2025}, SteP~\cite{step2023} and WebNavigator (ours) in \cref{table:action_space_comparison}.

\begin{table*}[!ht]
\centering
\caption{Comprehensive Comparison of Action Spaces across WebArena, BrowserGym (BG), AgentOccam (AO), SteP and WebNavigator.}
\label{table:action_space_comparison}
\small
\renewcommand{\arraystretch}{1.02}
\setlength{\tabcolsep}{1pt}
\begin{tabular}{l l c c c c >{\columncolor{blue!5}}c} 
\toprule
\textbf{Action Type} & \textbf{Description} & \textbf{WebArena} & \textbf{BG} & \textbf{AO} & \textbf{SteP} & \textbf{WebNavigator} \\
\midrule
\texttt{click(elem)} & Click at an element & \checkmark & \checkmark & \checkmark & \checkmark & \textbf{\checkmark} \\
\texttt{type(elem, text)} & Type to an element & \checkmark & \checkmark & \checkmark & \checkmark & \textbf{\checkmark} \\
\texttt{go\_back} & Visit the last URL & \checkmark & \checkmark & \checkmark & \checkmark & \textbf{\checkmark} \\
\texttt{note(content)} & Take notes & - & - & \checkmark & \checkmark & \textbf{\checkmark} \\
\texttt{stop(answer)} & Stop with an answer & - & - & \checkmark & \checkmark & \textbf{\checkmark} \\
\texttt{navigate(domain,query)} & \textbf{Teleport via multimodal retrieval} & - & - & - & - & \textbf{\checkmark} \\
\midrule
\texttt{noop} & Do nothing & \checkmark & \checkmark & - & - & - \\
\texttt{hover(elem)} & Hover on an element & \checkmark & \checkmark & - & \checkmark & - \\
\texttt{press(key\_comb)} & Press a key combination & \checkmark & \checkmark & - & \checkmark & - \\
\texttt{scroll(dir)} & Scroll up and down & \checkmark & \checkmark & - & \checkmark & - \\
\texttt{tab\_focus(index)} & Focus on the i-th tab & \checkmark & \checkmark & - & \checkmark & - \\
\texttt{new\_tab} & Open a new tab & \checkmark & \checkmark & - & \checkmark & - \\
\texttt{tab\_close} & Close current tab & \checkmark & \checkmark & - & \checkmark & - \\
\texttt{go\_forward} & Undo go\_back & \checkmark & \checkmark & - & \checkmark & - \\
\texttt{goto(URL)} & Go to URL & \checkmark & \checkmark & - & \checkmark & - \\
\texttt{dbclick(elem)} & Double-click at an element & - & \checkmark & - & - & - \\
\texttt{clear(elem)} & Clear the content & - & \checkmark & - & - & - \\
\texttt{focus(elem)} & Set the focus to an element & - & \checkmark & - & - & - \\
\texttt{select\_option(elem)} & Select elements within menu & - & \checkmark & - & - & - \\
\texttt{drag\_and\_drop(elem)} & Drag and drop element to another & - & \checkmark & - & - & - \\
\texttt{upload\_file(elem,file)} & Upload files & - & \checkmark & - & - & - \\
\texttt{send\_msg\_to\_user(text)} & Send a message to the user & - & \checkmark & - & - & - \\
\texttt{report\_infeasible(text)} & Report instructions are infeasible & - & \checkmark & - & - & - \\
\texttt{go\_home} & Go to home page & - & - & \checkmark & - & - \\
\texttt{branch(id, intent)} & Generate a new plan & - & - & \checkmark & - & - \\
\texttt{prune(id, reason)} & Restore to a previous plan & - & - & \checkmark & - & - \\
\texttt{find\_commits(query)} & Search commits in a project & - & - & - & \checkmark & - \\
\texttt{search\_issues(query)} & Search and filter issues & - & - & - & \checkmark & - \\
\texttt{create\_project(query)} & Create project and add members& - & - & - & \checkmark & - \\
\texttt{create\_group(query)} & Create group and add members& - & - & - & \checkmark & - \\
\texttt{find\_subreddit(query)} & Find specific or relevant subreddit & - & - & - & \checkmark & - \\
\texttt{find\_user(user\_name)} & Navigate to a user's page & - & - & - & \checkmark & - \\
\texttt{find\_review(query)} & Find reviews for a product & - & - & - & \checkmark & - \\
\texttt{find\_order(query)} & Find order by customer\/id& - & - & - & \checkmark & - \\
\texttt{search\_customer(query)} & Search customers by query details & - & - & - & \checkmark & - \\
\texttt{search\_order(question)} & Search orders to answer questions & - & - & - & \checkmark & - \\
\texttt{list\_products(query)} & List products matching query & - & - & - & \checkmark & - \\
\texttt{search\_reviews(query)} & Search reviews for answers & - & - & - & \checkmark & - \\
\texttt{find\_directions(query)} & Find directions between locations & - & - & - & \checkmark & - \\
\texttt{search\_nearest(query)} & Find places near a location & - & - & - & \checkmark & - \\
\midrule
\textbf{Action Space Size} & & \textbf{12} & \textbf{20} & \textbf{8} & \textbf{27} & \textbf{\textcolor{blue}{6}} \\
\bottomrule
\end{tabular}
\end{table*}

\subsection{Efficiency Analysis}
\begin{table}[ht]
\small
\centering
\caption{\;\; Average number of steps and average end-to-end\\ execution time per task on WebArena.}
\label{tab:efficiency}
\begin{tabular}{llccc}
\toprule
\multirow{2}{*}{Method} & \multirow{2}{*}{Model} & \multicolumn{2}{c}{Avg.\ Steps} & \multirow{2}{*}{Avg.\ Time} \\
\cmidrule(lr){3-4}
 & & All & Multi-site & \\
\midrule
\multirow{2}{*}{WebArena} & Claude-Sonnet-4  & 12.27 & 13.62 & 191.6s \\
                          & Gemini-2.5-Pro   & 9.15  & 10.98 & 128.8s \\
\midrule
\multirow{2}{*}{AgentOccam} & Claude-Sonnet-4 & 10.14 & 14.56 & 643.3s \\
                            & Gemini-2.5-Pro  & 10.10 & 14.96 & 773.6s \\
\midrule
\multirow{2}{*}{WebNavigator} & Claude-Sonnet-4 & 8.93 & 11.75 & 635.1s \\
                              & Gemini-2.5-Pro  & 8.98 & 10.75 & 753.5s \\
\bottomrule
\end{tabular}
\end{table}

As shown in~\cref{tab:efficiency}, WebNavigator achieves superior efficiency in both trajectory length and online execution time compared to the previous state-of-the-art. WebNavigator with Claude-Sonnet-4 and Gemini-2.5-Pro requires an average of 8.93 and 8.98 steps per task, compared to 10.14 and 10.10 for AgentOccam. The efficiency gains are most pronounced on Multi-site tasks, where Topological Blindness is most severe. WebNavigator with Claude-Sonnet-4 averages 11.75 steps on Multi-site tasks versus 14.56 for AgentOccam, a 19.3\% reduction. With Gemini-2.5-Pro, this reduction reaches 28.1\%, from 14.96 steps to 10.75. These results confirm that the \texttt{navigate} action produces shorter trajectories by teleporting agents directly to task-relevant pages, bypassing the redundant exploratory steps that reactive agents must perform.
Regarding online execution time, WebNavigator remains competitive with AgentOccam under identical inference conditions. With Claude-Sonnet-4, WebNavigator completes tasks in 635.1 seconds on average compared to 643.3 seconds for AgentOccam. With Gemini-2.5-Pro, WebNavigator requires 753.5 seconds versus 773.6 seconds. These results demonstrate that the Retrieve-Reason-Teleport workflow introduces no additional online overhead relative to the baseline, while delivering substantially higher task success rates.

\subsection{None Returns Do Not Indicate Task Failure}
\label{sec:none_returns}
When the Selector returns None, indicating that no candidate sufficiently matches the navigation intent, the Teleport stage is bypassed and the agent continues task execution from the current page via low-level actions. To quantify how often this occurs and whether it leads to task failure, we measure the None return frequency of the Selector and the subsequent task completion rate across all WebArena tasks for four backbone models.
As shown in~\cref{tab:none_returns}, None returns are rare across all configurations, with GPT-4o exhibiting the lowest rate at 3.6\% and no model exceeding 9.1\%. 
More importantly, between 43.6\% and 62.2\% of None-triggered tasks are still completed successfully across backbone models. These results confirm that None returns do not indicate task failure, and the fallback mechanism ensures WebNavigator degrades gracefully under retrieval uncertainty.

\begin{table}[h]
  \centering
  \small
  \caption{None return frequency and task success rate on None-triggered tasks across all WebArena tasks.}
  \label{tab:none_returns}
  \begin{tabular}{lcc}
  \toprule
  \textbf{Model} & \textbf{None Rate} & \textbf{Fallback SR} \\
  \midrule
  Qwen3-VL-32B    & 6.8\% & 43.6\% \\
  GPT-4o          & 3.6\% & 51.7\% \\
  Claude-Sonnet-4 & 5.3\% & 58.1\% \\
  Gemini-2.5-Pro  & 9.1\% & 62.2\% \\
  \bottomrule
\end{tabular}
\end{table}

\section{Additional Baselines}
To ensure a fair comparison across backbone models, we extend the evaluation of WebArena and 
AgentOccam baselines beyond GPT-4o to include Qwen3-VL-32B-Instruct, Claude-Sonnet-4, and 
Gemini-2.5-Pro. As shown in Table~\ref{tab:additional_baselines}, WebNavigator consistently outperforms both 
baselines across all backbone models by a substantial margin. These results confirm that WebNavigator's improvements are attributable to the framework design 
rather than the choice of backbone model.
\begin{table}[h]
\centering
\caption{WebArena success rate (\%) across additional models.}
\label{tab:additional_baselines}
\begin{tabular}{lccc}
\toprule
\textbf{Method} & \textbf{Qwen3-VL-32B-Instruct} & \textbf{Claude-Sonnet-4} & 
\textbf{Gemini-2.5-Pro} \\
\midrule
WebArena    & 19.1 & 19.8 & 15.9 \\
AgentOccam  & 34.2 & 49.3 & 50.4 \\
WebNavigator (Ours) & \textbf{47.8} & \textbf{57.1} & \textbf{63.3} \\
\bottomrule
\end{tabular}
\end{table}

\section{Additional Evaluation on WebVoyager}
\label{sec:webvoyager_results}

WebVoyager \cite{webvoyager2024} evaluates end-to-end web navigation on live real-world websites. We reproduce the official WebVoyager agent and evaluate both methods with GPT-4o on the same tasks from five WebVoyager domains. WebNavigator outperforms the reproduced WebVoyager agent on four of the five evaluated domains and matches it on Apple, providing complementary evidence of generalization beyond WebArena.

\begin{table}[htbp]
\centering
\small
\setlength{\tabcolsep}{5pt}
\caption{Success rate (\%) on five WebVoyager domains using GPT-4o.}
\label{tab:webvoyager_results}
\begin{tabular}{lccccc}
\toprule
\textbf{Method} & \textbf{Allrecipes} & \textbf{Apple} & \textbf{Arxiv} & \textbf{BBC News} & \shortstack{\textbf{Cambridge} \\ \textbf{Dictionary}} \\
\midrule
WebVoyager  & 44.4 & \textbf{51.2} & 34.9 & 47.6 & 69.8 \\
WebNavigator & \textbf{77.8} & \textbf{51.2} & \textbf{62.8} & \textbf{66.7} & \textbf{90.7} \\
\bottomrule
\end{tabular}
\end{table}

\section{Detailed Query-Level Selector Analysis}
\label{sec:selector_details}

\begin{wraptable}{r}{0.44\textwidth}
\centering
\small
\setlength{\tabcolsep}{3pt}
\caption{Query annotation statistics.}
\label{tab:selector_query_stats}
\begin{tabular}{lcc}
\toprule
\textbf{Backbone} & \textbf{Queries} & \textbf{Valid / No Target} \\
\midrule
GPT-4o          & 91 & 74 / 17 \\
Claude-Sonnet-4 & 61 & 49 / 12 \\
Gemini-2.5-Pro  & 50 & 47 / 3  \\
\bottomrule
\end{tabular}
\end{wraptable}

For each navigation query, annotators inspect the corresponding domain-specific Interaction Graph and determine whether a valid target page exists. The resulting annotation statistics are summarized in \cref{tab:selector_query_stats}.
Retriever Top-1 Accuracy measures whether the highest-ranked retrieved page is a valid target. Target Recall@30 measures whether at least one valid target appears among the top-30 retrieved candidates. Both metrics are computed over queries for which a valid graph target exists. Selector Accuracy is computed over all queries and counts both valid target selections and correct \texttt{None} predictions.

\section{Heuristic Auto-Exploration Engine vs.\ LLM-based Exploration Engine: A Controlled Comparison}
\label{sec:auto_vs_llm}

The Heuristic Auto-Exploration Engine captures the topological skeleton of a website via click-based navigation. A natural question is whether incorporating type-based interactions would uncover additional functional entry points that click-based exploration misses. To answer this, we compared our click-based Adaptive BFS against a variant that additionally invokes an LLM to generate type and press actions on input elements, and measured the functional entry point coverage of both methods.

\paragraph{Experimental Setup.}
A functional entry point is a page state from which a task can be executed directly, requiring no further cross-page navigation. Multiple tasks can share the same functional entry point, so we first establish the ground-truth set by human annotation. Three annotators with doctoral-level or senior web development expertise independently labeled the functional entry point for each of the 106 tasks in the WebArena Reddit domain. The final label was determined by majority vote; no task produced three divergent labels, indicating reliable annotation agreement. The annotation identified 16 unique functional entry points across the 106 tasks.

\begin{table}[ht]
\centering
\caption{Comparison of Heuristic Auto-Exploration Engine and LLM-based Exploration Engine on the WebArena Reddit domain. Coverage reports functional entry point discovery out of 16 human-annotated ground-truth points.}
\label{tab:exploration_comparison}
\begin{tabular}{lccc}
\toprule
\textbf{Method} & \textbf{Depth} & \textbf{Total Nodes} & \textbf{Coverage} \\

\midrule
Heuristic Auto-Exploration Engine       & 2 & 48  & 8/16  \\
Heuristic Auto-Exploration Engine       & 3 & 225 & 16/16 \\
LLM-based Exploration Engine  & 2 & 55  & 8/16  \\
LLM-based Exploration Engine  & 3 & 241 & 16/16 \\
\bottomrule
\end{tabular}
\end{table}

We compare two exploration strategies at depths 2 and 3. The \textbf{Heuristic Auto-Exploration Engine} uses the click-based Adaptive BFS described in~\cref{sec:phase1}. The \textbf{LLM-based Exploration Engine} extends the same Adaptive BFS backbone by invoking Gemini-2.5-Pro on input-capable elements. Specifically, when the engine encounters an element with a WAI-ARIA input role (\texttt{searchbox}, \texttt{textbox}, or \texttt{combobox}), it provides the current page URL and the element's attributes to generate a contextually plausible \texttt{type} action followed by a \texttt{press} of the Enter key (full prompt provided in Appendix~\ref{sec:llm_explore_prompt}).\footnote{Web Accessibility Initiative Accessible Rich Internet Applications (WAI-ARIA) is a W3C standard that defines a set of roles and attributes to make web content more accessible. \url{https://www.w3.org/TR/wai-aria/}} 
All other interactive roles are handled identically to the Heuristic Auto-Exploration Engine via click-based interactions. The two methods share an identical Adaptive BFS backbone, with the sole difference being whether type-based interactions are applied, ensuring that any additional nodes discovered by the LLM-based Exploration Engine are attributable solely to type-based interactions.

\begin{figure*}[h]
  \centering
  \begin{minipage}{0.3\textwidth}
    \centering
    \includegraphics[width=\linewidth]{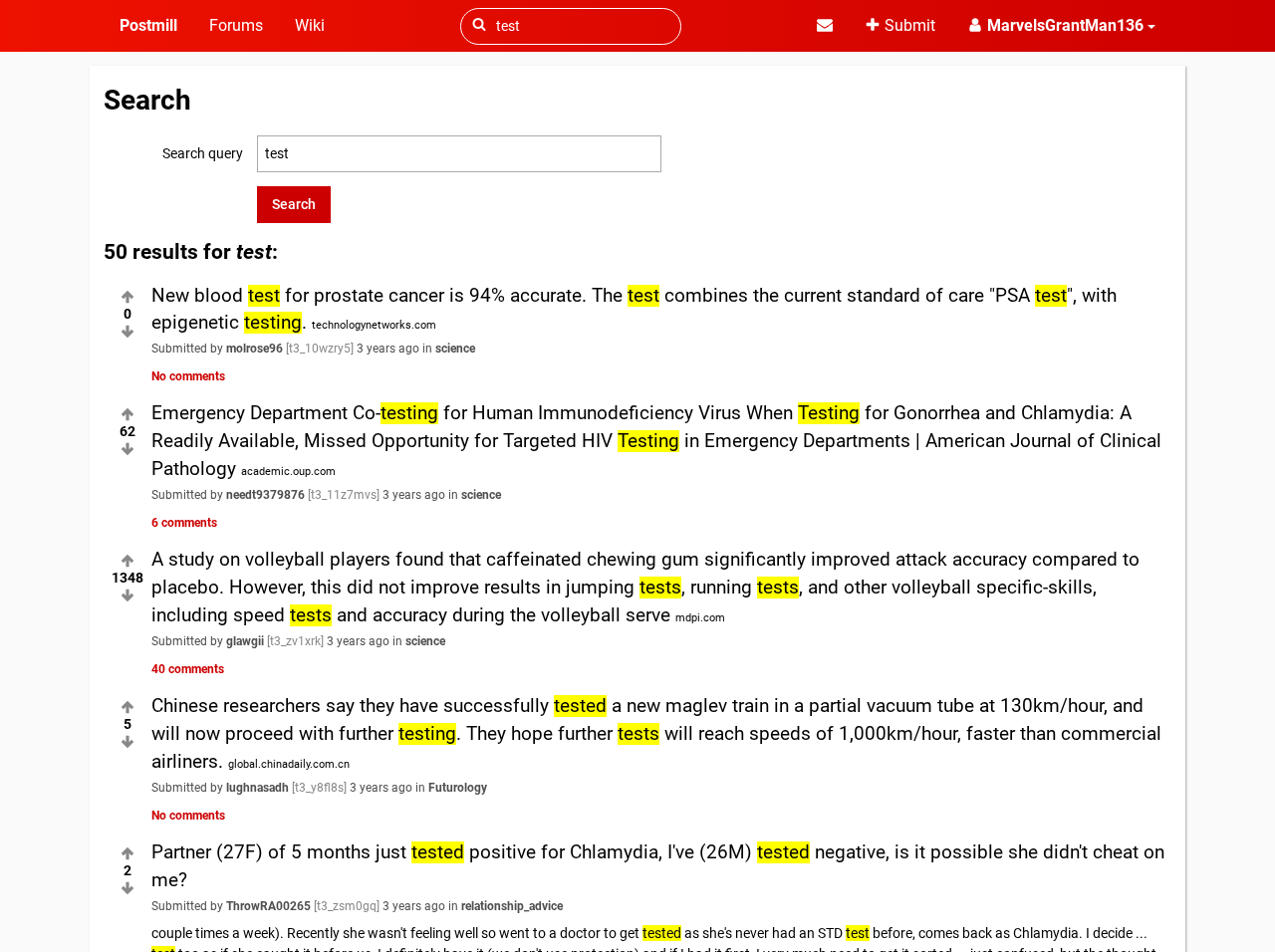}
  \end{minipage}
  \begin{minipage}{0.3\textwidth}
    \centering
    \includegraphics[width=\linewidth]{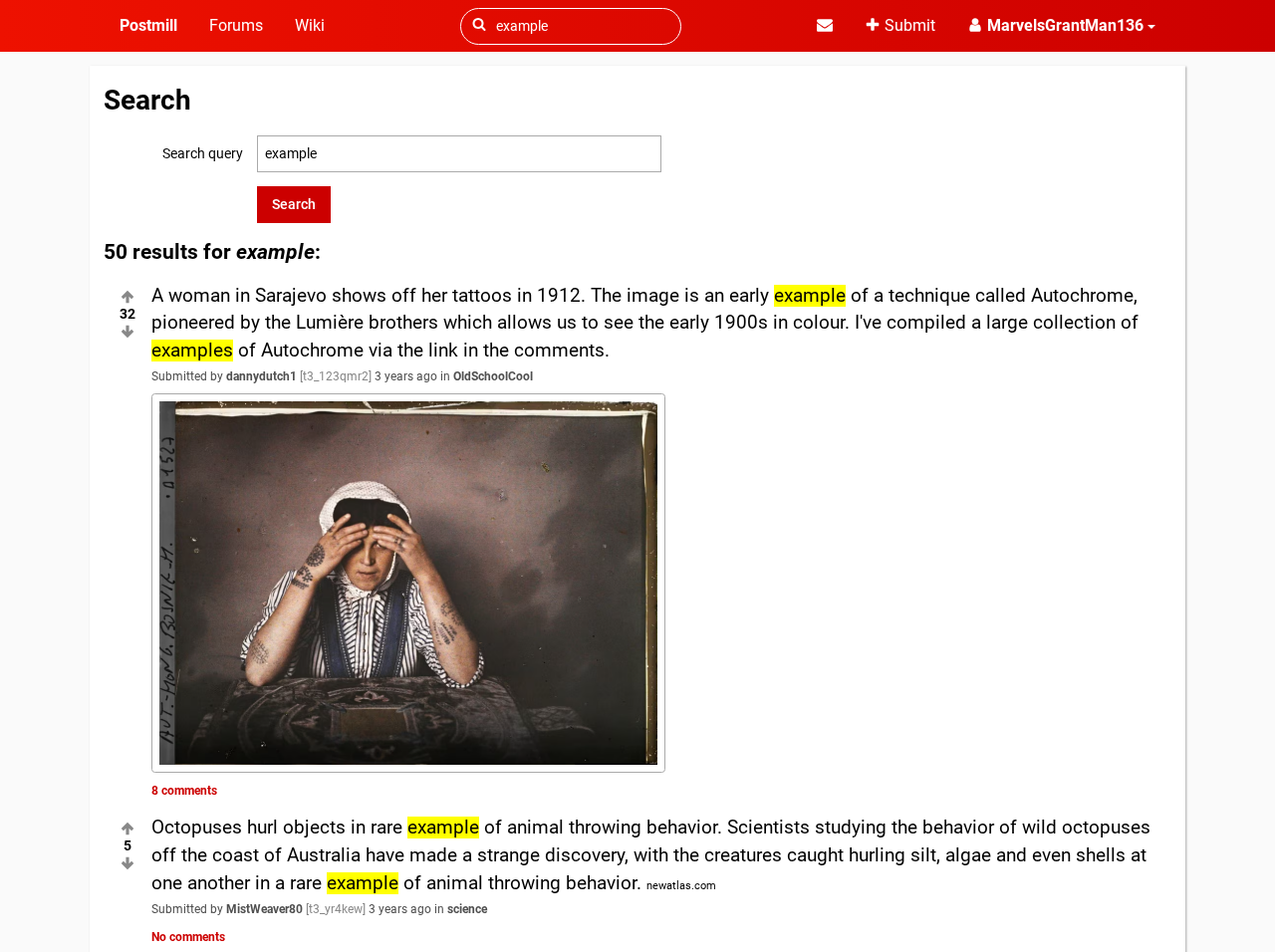}
  \end{minipage}
  \begin{minipage}{0.3\textwidth}
    \centering
    \includegraphics[width=\linewidth]{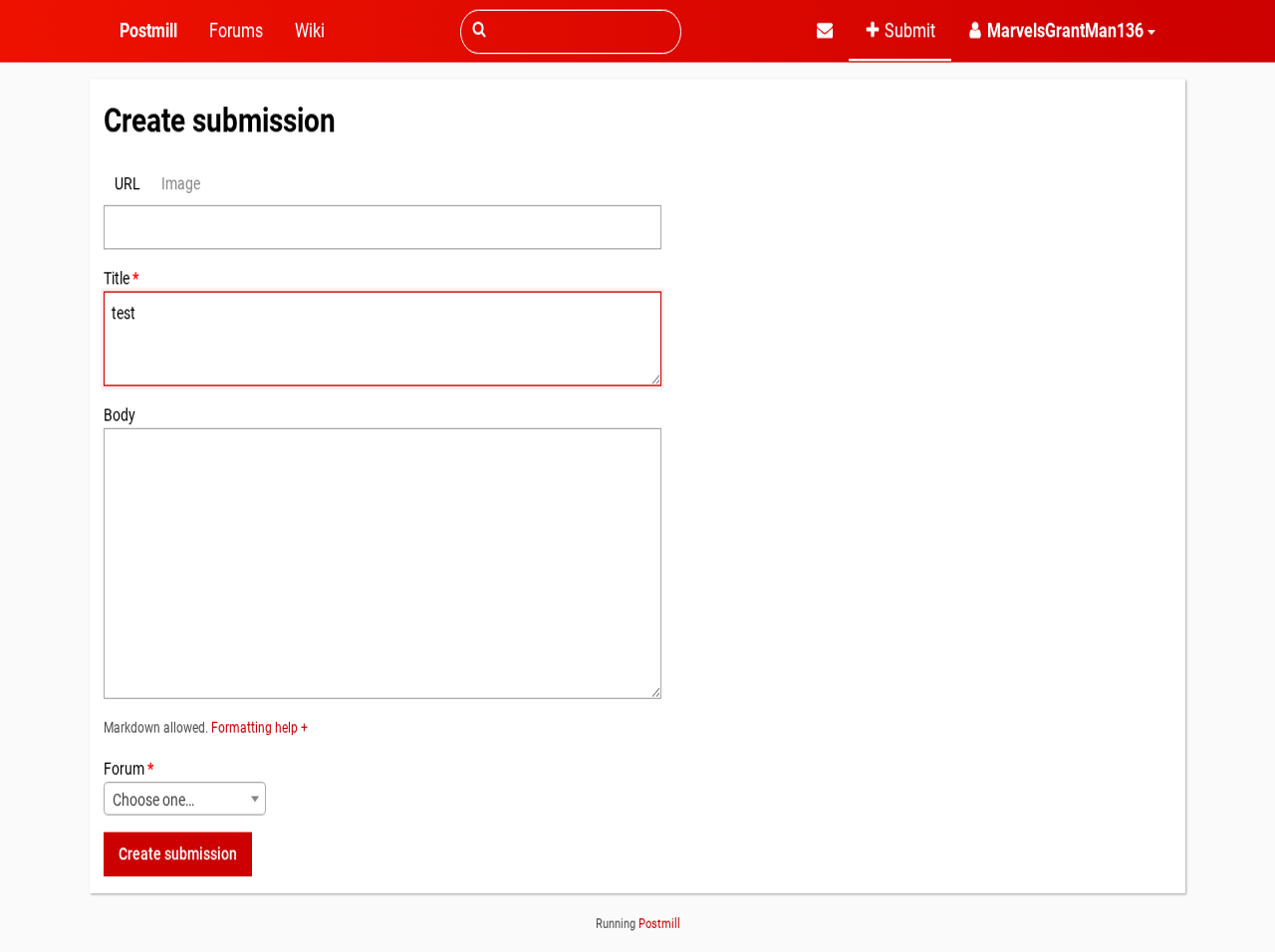}
  \end{minipage}

  \vspace{0.5em}

  \begin{minipage}{0.3\textwidth}
    \centering
    \includegraphics[width=\linewidth]{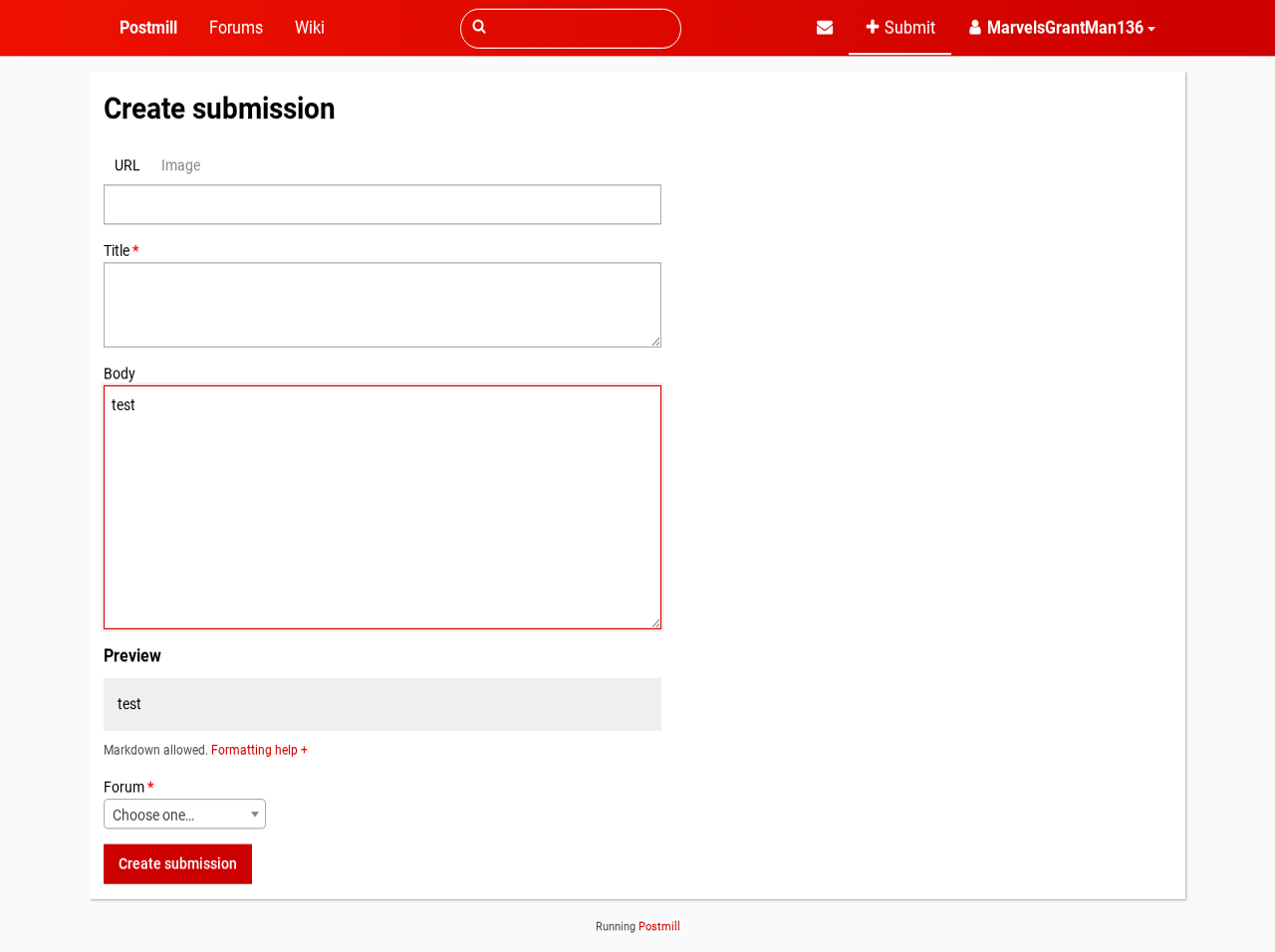}
  \end{minipage}
  \begin{minipage}{0.3\textwidth}
    \centering
    \includegraphics[width=\linewidth]{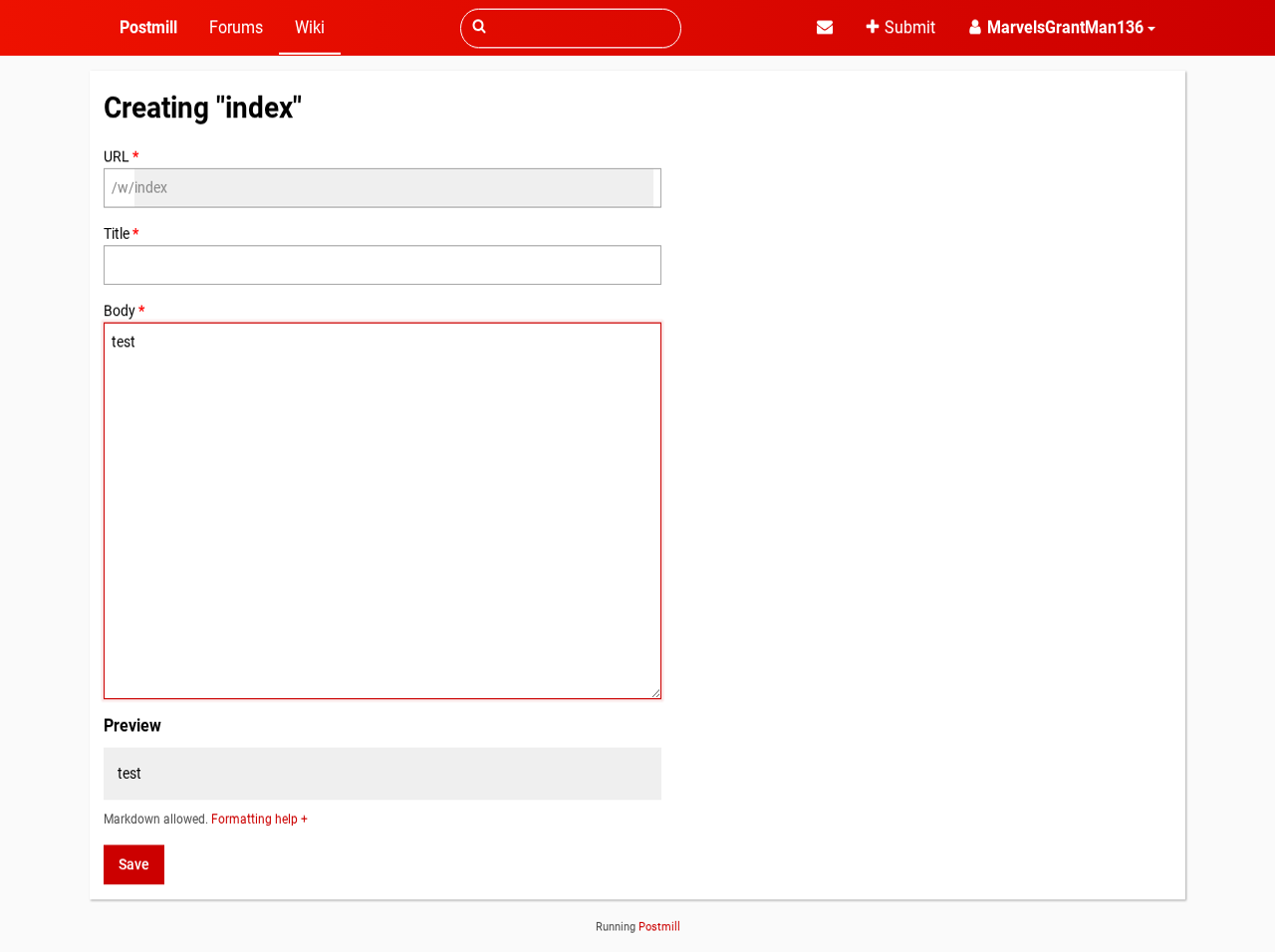}
  \end{minipage}
  \begin{minipage}{0.3\textwidth}
    \centering
    \includegraphics[width=\linewidth]{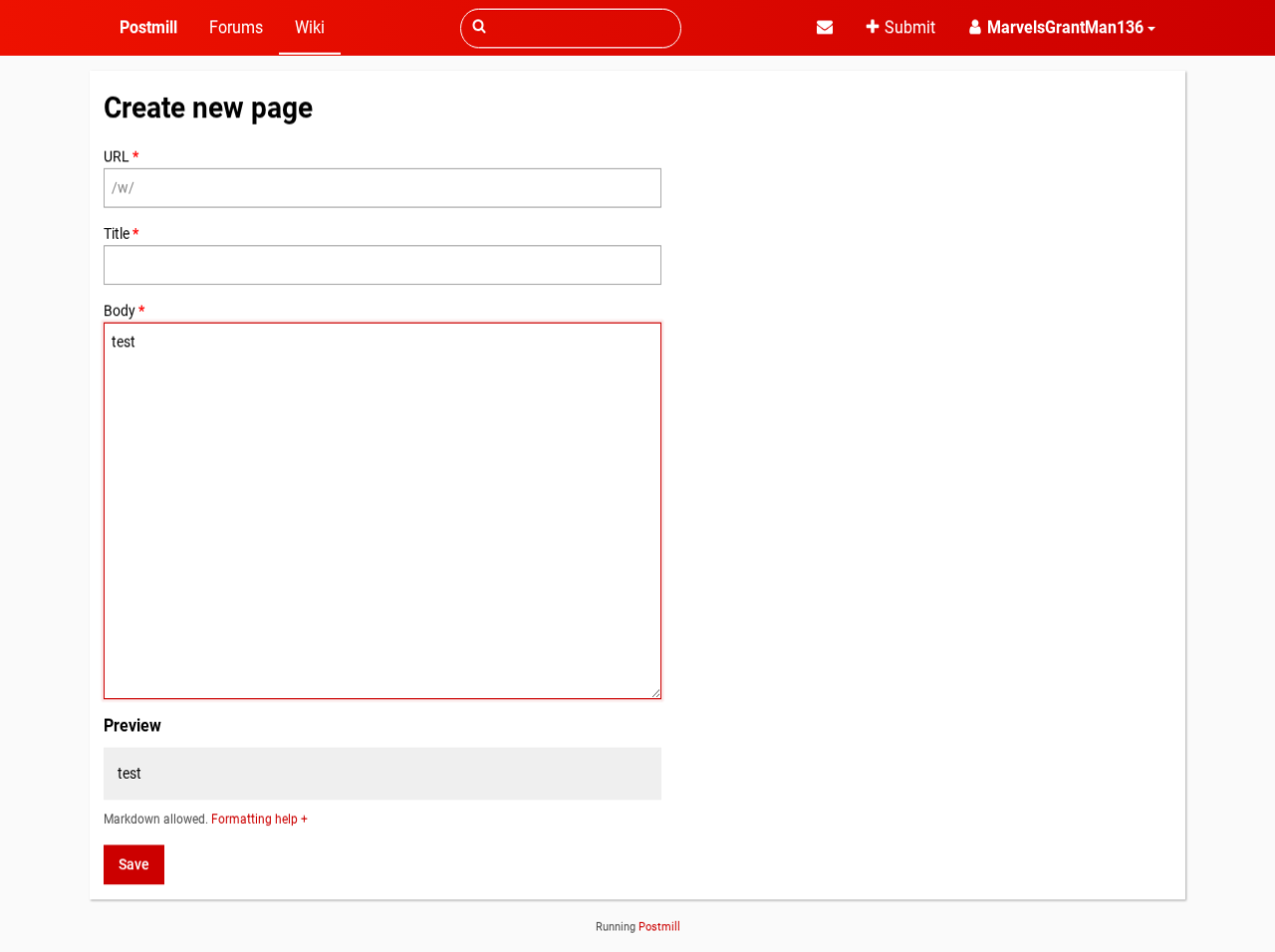}
  \end{minipage}

  \vspace{0.5em}

  \begin{minipage}{0.3\textwidth}
    \centering
    \includegraphics[width=\linewidth]{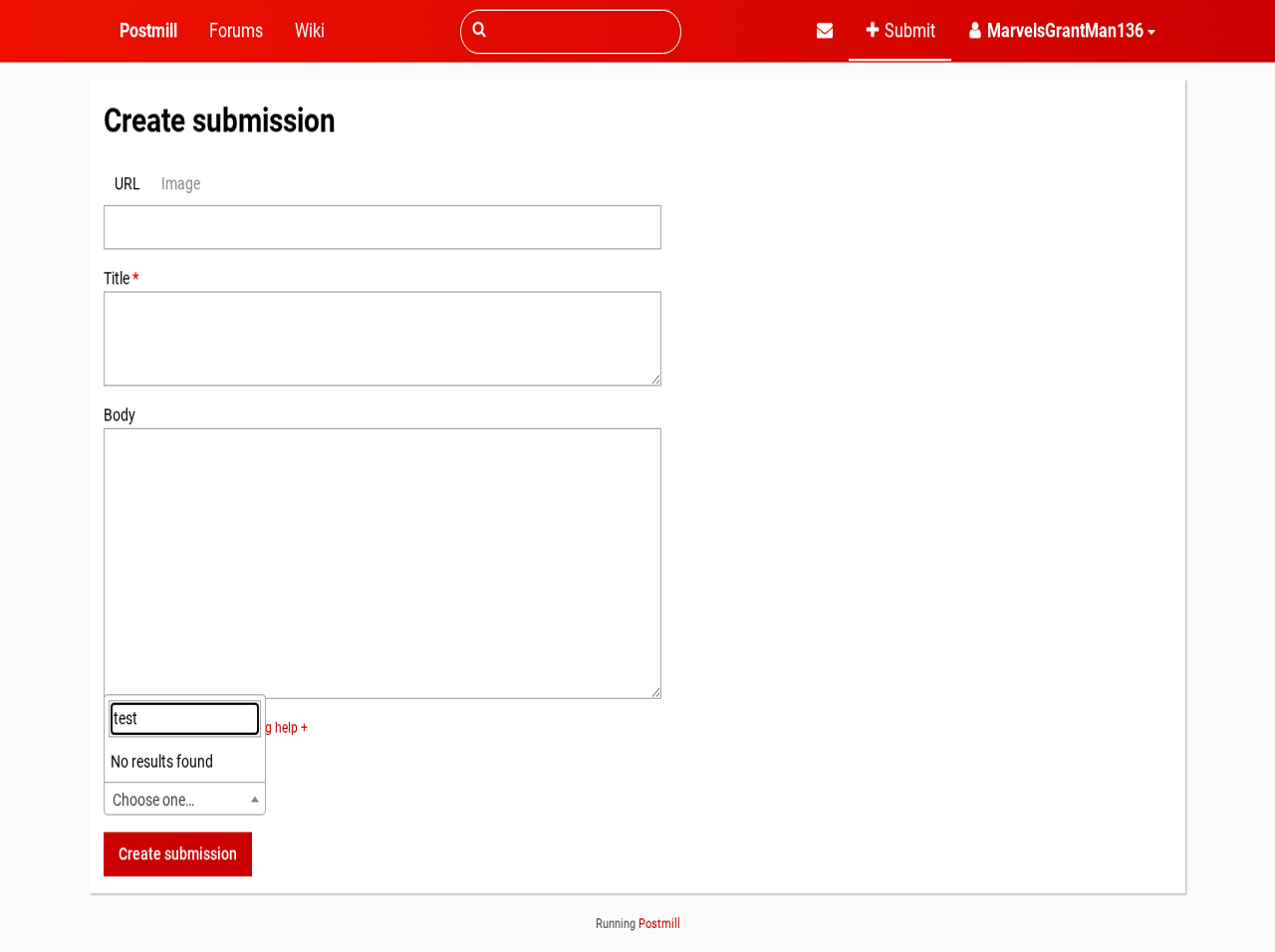}
  \end{minipage}
  \begin{minipage}{0.3\textwidth}
    \centering
    \includegraphics[width=\linewidth]{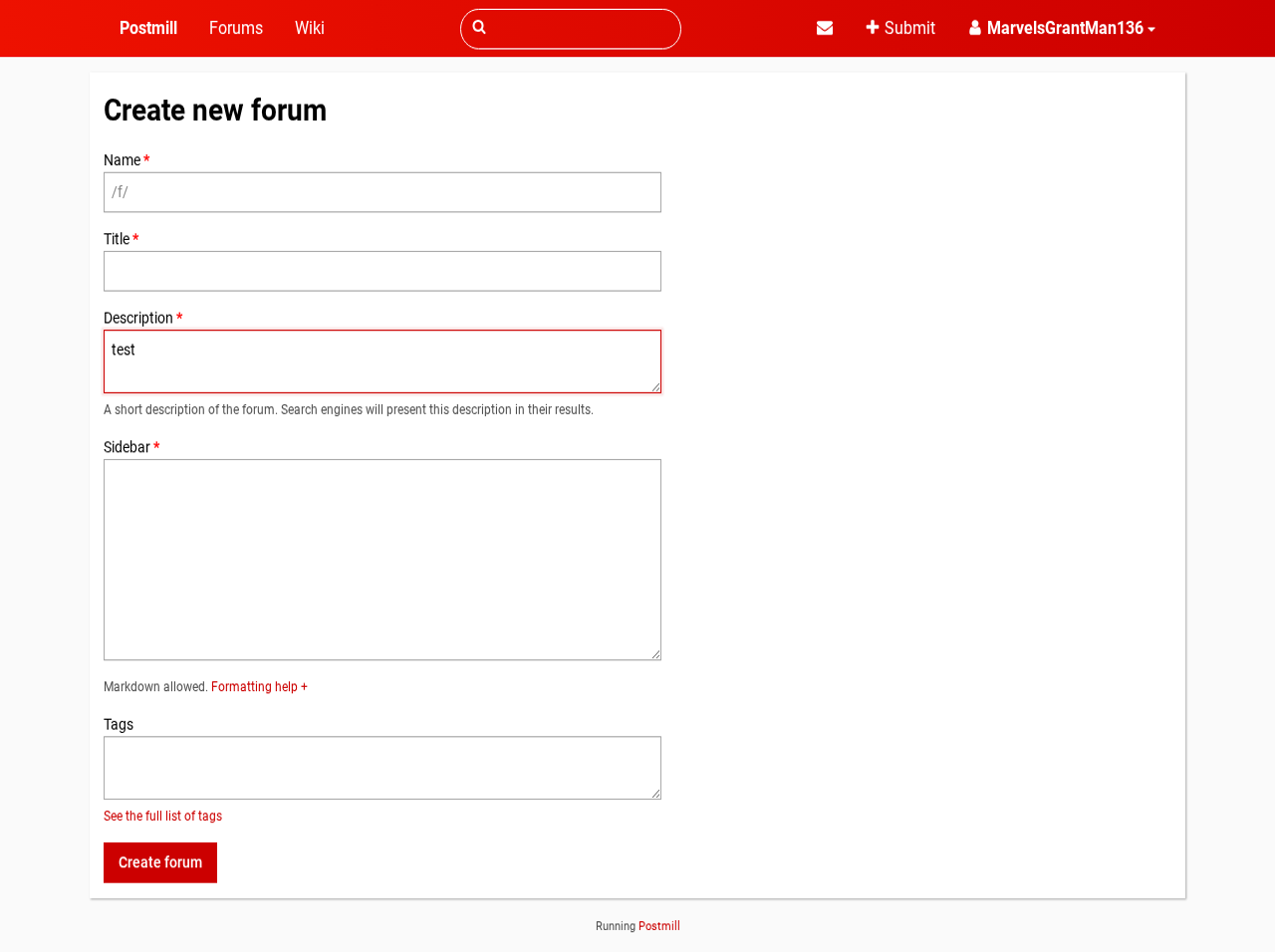}
  \end{minipage}
  \begin{minipage}{0.3\textwidth}
    \centering
    \includegraphics[width=\linewidth]{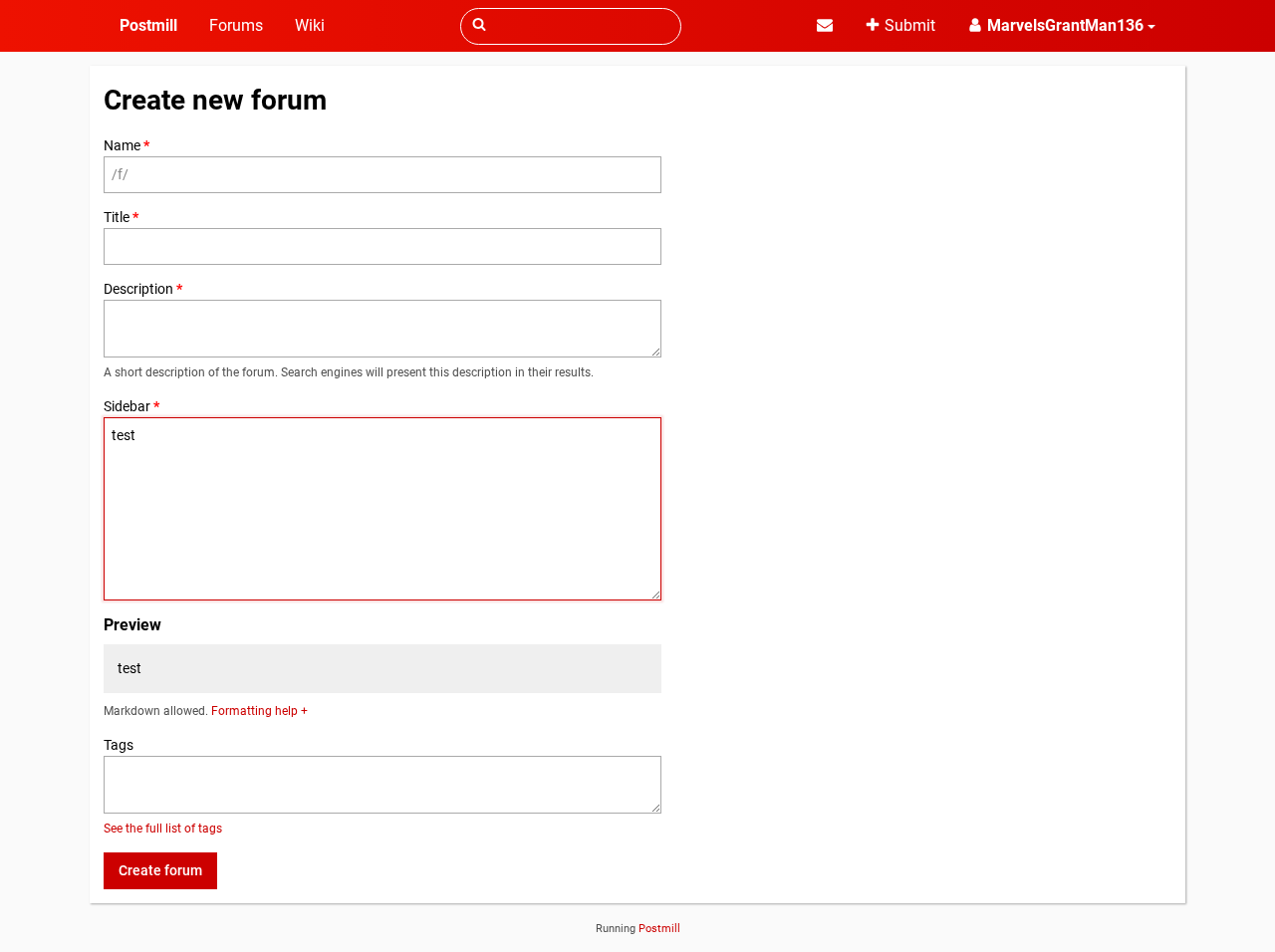}
  \end{minipage}

  \caption{Representative page states discovered exclusively by the LLM-based Exploration Engine via type-based interactions at depth 3 on the WebArena Reddit domain, including search result pages and form-filling interfaces.}
  \label{fig:type_triggered}
\end{figure*}

\paragraph{Results.}
As shown in Table~\ref{tab:exploration_comparison}, both methods achieve identical coverage at each depth: 8/16 at depth 2 and 16/16 at depth 3, confirming that click-based exploration is sufficient to achieve complete functional coverage of the Reddit domain. The LLM-based Exploration Engine discovers more nodes than the Heuristic Auto-Exploration Engine, increasing from 48 to 55 at depth 2 and from 225 to 241 at depth 3, yet none of these additional nodes correspond to any of the 16 functional entry points.
Manual inspection confirms that all additional nodes are type-triggered states whose functional entry points are already present in the click-based graph. Figure~\ref{fig:type_triggered} presents representative examples of such type-triggered states, all of which are search result pages or form-filling interfaces that introduce no new navigational pathways beyond those already captured by the Heuristic Auto-Exploration Engine. These results confirm that type-based interactions are task-dependent by nature and are therefore handled in the online Local Execution Stage rather than the offline exploration stage.

\section{End-to-End Cost Accounting}
\label{sec:end2end_cost}
A more detailed end-to-end cost analysis further improves the completeness of our evaluation. We provide a stage-wise accounting of graph exploration, embedding and indexing, graph maintenance, and online inference. Because offline graph costs are reusable across tasks, while online costs are incurred for every task, we report them separately.

\paragraph{Offline Graph Construction and Indexing.}
Appendix~\ref{subsec:graph_stats} and Table~\ref{tab:graph_stats} in the paper already report graph exploration time for all WebArena domains, including all browser interactions performed by the exploration engine. We additionally measure the complete embedding-and-indexing stage from the first embedding request until the persistent index is written to disk.

\begin{table}[htbp]
\centering
\small
\setlength{\tabcolsep}{4.5pt}
\caption{Offline graph construction and indexing cost across WebArena domains.}
\label{tab:offline_cost}
\begin{tabular}{l c c c c c}
\toprule
\textbf{Domain} & \textbf{Depth} & \textbf{\shortstack{Exploration\\(min)}} & \textbf{\shortstack{Embedding \&\\Indexing (min)}} & \textbf{\shortstack{Index Size\\(MB)}} & \textbf{\shortstack{API Cost\\(USD)}} \\
\midrule
Map      & 1 &   2.1 &  0.31 &  16.18 & \$0.01 \\
CMS      & 2 &  23.1 &  1.99 & 103.28 & \$0.17 \\
Shopping & 2 & 122.1 & 12.96 & 470.02 & \$1.26 \\
GitLab   & 2 & 316.8 & 21.75 & 677.67 & \$1.13 \\
Reddit   & 3 &  31.1 &  3.82 & 164.62 & \$0.38 \\
\bottomrule
\end{tabular}
\end{table}

Graph exploration requires commodity CPU hardware without GPU or LLM calls. As shown in Table~\ref{tab:offline_cost}, the largest domain, GitLab, takes 316.8 minutes and costs approximately USD 0.90 on an AWS c7i.xlarge instance. Embedding and indexing take 0.31--21.75 minutes, cost at most USD 1.26 per domain, and require at most 677.67 MB of persistent storage. These website-level costs are shared across subsequent tasks.

\paragraph{Online \texttt{navigate} Cost.}
To complete the online cost accounting, we instrument each stage inside the high-level navigate action on the full WebArena benchmark using Gemini-2.5-Pro under the same configuration as our main experiments. All values are averaged over the complete task set.
Retrieve is measured from submitting the navigation query until all candidates are returned. Reason is measured from submitting the candidate pages until the Selector output is received. Teleport is measured from shortest-path computation until all underlying browser actions have completed and the target page has loaded.

\begin{table}[htbp]
\centering
\small
\caption{Online cost breakdown of \textit{navigate}.}
\label{tab:navigate_cost_breakdown}
\begin{tabular}{lcc}
\toprule
\textbf{Component} & \textbf{Avg. Time / Task} & \textbf{Avg. Tokens / Task} \\
\midrule
Retrieve           & 2.2 s                     & 83 input                    \\
Reason             & 11.1 s                    & 6,169 input $+$ 1,800 output \\
Teleport           & 2.3 s                     & ---                         \\
\midrule
\textit{navigate} workflow & 15.6 s            & 8,052 total                 \\
\bottomrule
\end{tabular}
\end{table}

\paragraph{End-to-End Comparison.}
To contextualize the stage-level costs, we also measure end-to-end execution time and total token consumption under the same Gemini-2.5-Pro WebArena setting, as shown in Table~\ref{tab:end_to_end_cost}.

\begin{table}[htbp]
\centering
\small
\caption{End-to-end online cost and performance on the WebArena benchmark.}
\label{tab:end_to_end_cost}
\begin{tabular}{llccc}
\toprule
\textbf{Method} & \textbf{Model} & \textbf{Avg. Time / Task} & \textbf{Avg. Tokens / Task} & \textbf{SR} \\
\midrule
AgentOccam   & Gemini-2.5-Pro & 773.6 s           & 67,351            & 50.40\% \\
WebNavigator & Gemini-2.5-Pro & \textbf{753.5 s}  & 123,684           & \textbf{63.30\%} \\
\bottomrule
\end{tabular}
\end{table}

The complete navigate workflow requires 15.6 seconds per task, accounting for 2.1\% of WebNavigator's end-to-end execution time. The Selector consumes 7,969 tokens per task, representing 6.4\% of the end-to-end model tokens. Teleport requires only 2.3 seconds and no model tokens. These measurements quantify the operations encapsulated by this high-level action and show that they introduce limited online overhead.

Compared with AgentOccam under the same backbone, WebNavigator improves task success from 50.4\% to 63.3\% while reducing average execution time from 773.6 to 753.5 seconds. WebNavigator uses more end-to-end model tokens, and we report this trade-off explicitly.
Overall, WebNavigator delivers a 12.9-point success improvement over AgentOccam with slightly lower execution latency, while the additional online overhead introduced by navigate remains small.

\section{The Contribution of the Interaction Graph}
\label{sec:graph_contribution}
To rigorously isolate the effect of topology awareness, we used the Gemini-2.5-Pro setting of \cref{tab:exp1_results} and replaced the Interaction Graph with an empty graph containing no nodes. All other settings remained fixed. The Retriever returned no candidates and the Selector returned \texttt{None}, while low-level actions remained unchanged. We then reran all 514 tasks solved by full WebNavigator, of which the empty-graph variant failed on 95.

Because this evaluation is conditioned on full-system success, it measures regressions among previously successful tasks rather than the overall success rate of the empty-graph variant. We therefore manually classified them to identify the source of the performance degradation. As shown in \cref{tab:empty_graph_error_analysis}, 37 of the 95 regressions are attributed to Topological Blindness, accounting for 38.9\%. This controlled result shows that removing topology awareness from WebNavigator directly reintroduces Topological Blindness.

\begin{table}[htbp]
\centering
\small
\caption{Empty-graph error analysis and classification of regressed tasks.}
\label{tab:empty_graph_error_analysis}
\begin{tabular}{lp{6.2cm}cc}
\toprule
\textbf{Failure Type} & \textbf{Definition} & \textbf{Count} & \textbf{Share} \\
\midrule
Topological Blindness & Target page never located during online exploration & 37 & 38.90\% \\
Execution Error & Target page reached but local execution failed & 39 & 41.10\% \\
Step Limit Exhaustion & Step budget exhausted & 18 & 18.90\% \\
Annotation Error & Incorrect ground truth & 1 & 1.10\% \\
\bottomrule
\end{tabular}
\end{table}

Beyond this binary ablation of graph availability, the Knowledge Completeness experiment in \cref{sec:impact_of_knowledge_completeness} further examines how graph coverage affects performance. With GPT-4o and all online components fixed, increasing exploration depth from 1 to 4 raises the success rate from 63.2\% to 75.5\%. Together, the empty-graph ablation measures the contribution of graph availability, while the depth experiment measures the effect of graph coverage.

\section{Prompt}
\subsection{System Prompt}
The system prompt defines the behavioral logic of WebNavigator via three integrated components: the \textbf{Explore-Act Model}, which distinguishes between exploration and local action phases; the \textbf{Output Specifications} detailed in Appendix~\ref{sec:output_specifications}, which facilitate structured Chain-of-Thought reasoning; and the \textbf{Action Space Definition} detailed in Appendix~\ref{sec:action_definition}, which specifies the executable actions available to the agent.

\begin{promptbox}[System Prompt]{sys}

You are an AI assistant performing tasks on a web browser. You will be provided with task objective, current step, web page observations, and other relevant information. You need to issue an action for this step.
\# Web Navigation Philosophy: The Explore-Act Model
Your operation must follow a strict two-phase model: the Exploration Phase and the Low-Level Action Phase. Any task can be composed of an Exploration Phase and a Low-Level Action Phase.
Covers Scenario 1: Explore -> low-level action
Covers Scenario 2: Explore -> low-level actions -> Explore -> low-level actions (can be repeated N times)
Covers Scenario 3: Explore (Domain A) -> low-level actions (Domain A) -> Explore (Domain B) -> low-level actions (Domain B) (can be repeated N times)
1. Exploration Phase (Jumping to the Right Place)
An Exploration Phase begins when you need to find a new starting point for a task or sub-task, and the current page is inefficient for direct navigation.
You MUST use optimal\_navigate tool to initiate any Exploration Phase, which occurs when:
A) Starting a New Task: You are at the beginning of a mission, and the current page (e.g., start page or homepage) is not your target workspace.
Covers Scenario 1: Explore -> low-level action
B) Transitioning to a New Sub-Task (Single Domain): You have completed one part of a task and must now navigate to a completely different functional area of the same website to begin the next part.
Example: After successfully adding a new product in shopping\_admin, your next goal is to create a marketing campaign for it. You would use this tool to jump from the Product Catalog to the Marketing Promotions page.
Covers Scenario 2: Explore -> low-level actions -> Explore -> low-level actions
C) Transitioning to a New Sub-Task (Cross-Domain): You have completed a sub-task on one website and now need to switch to a different domain to continue the mission.
Example: After finding a project name on reddit, you must switch to gitlab to find its code repository. This domain switch marks the beginning of a new Exploration Phase.
Covers Scenario 3: Explore (Domain A) -> low-level actions (Domain A)-> Explore (Domain B) -> low-level actions (Domain B)
2. Low-Level Action Phase (Interacting with the Page)
This phase involves direct interaction with the elements on the current page to achieve a specific goal.
A) You are interacting with page elements: This includes filling out forms, clicking buttons within a modal, typing into search bars, selecting from dropdowns, or modifying data on the current page.
B) You are already on the correct page: If the target is already visible and reachable via one clicks, perform those clicks manually.   
optimal\_navigate is STRICTLY FORBIDDEN during the Low-Level Action Phase.
Critical Rule: 
Once you've used optimal\_navigate to reach a page, this means the Exploration Phase is over, and the next step must be the Low-Level Action Phase. DO NOT use it again until you complete the current sub-task, or you will be severely punished!
Each task must invoke optimal\_navigate at least once. Repeated calls to optimal\_navigate are prohibited.
STRICT RULE: Use stop exclusively for final answers. Never use it for intermediate notes; use the note action for that purpose.
Generate the response in the following format:
\texttt{\{output specifications\}}
CRITICAL FORMATTING RULE - VIOLATION WILL RESULT IN IMMEDIATE TERMINATION:
You are FORBIDDEN from generating ANY text surrounded by double asterisks (**) such as **ACTION:**, **REASON:**, **PLAN:**, **OBJECTIVE:**, or similar bold markdown formatting.
This includes but is not limited to: **ACTION:**, **REASON:**, **PLAN:**, **OBJECTIVE:**, **ANALYSIS:**, **DECISION:**, **RESPONSE:**, **NEXT STEP:**, or any other capitalized words in double asterisks.
If you generate even ONE instance of this forbidden format, your response will be rejected and you will receive severe punishment.
Instead, always output content directly without any markdown formatting, headers, or structural markers.
You are ONLY allowed to use the following action commands. Strictly adheres to the given format. Only issue one single action.
\texttt{\{Action Space Definition\}}
\end{promptbox}

\subsection{Output Specifications}
\label{sec:output_specifications}
This section dictates the structured format for the response of agent, requiring explicit fields for context retention, history summarization, and step-by-step reasoning. It enforces a chain-of-thought process to validate current observations and mode analysis before the final action is generated.
\begin{promptbox}[Output Specifications]{specifications}
REPEAT FINAL OBJECTIVE:
Repeat the original objective to maintain context clarity across multiple react iterations.
Output the original objective exactly as it was stated at the beginning, ensuring all details and requirements are preserved. This helps prevent context degradation and keeps the LLM focused on the original goal in long-running interactions.

~
EXTRACT CONSTRAINTS:
Extract and classify ALL specific requirements from the objective in EXTRACT CONSTRAINTS section. Be exhaustive - identify every constraint, filter, or requirement mentioned or implied.
- time constraints: Any temporal requirements (e.g. years: ``2022'', dates: ``March 15'', periods: ``last month'', ``Q1'', ``this week'', ranges: ``between 2021-2023'')
- quantity constraints: Rankings (e.g. ``top-1'', ``most popular''), limits (e.g. ``at least 5'', ``maximum 10''), comparisons (e.g. ``cheaper than \$70'', ``higher rated than 6.5'')
- entity constraints: Specific names, IDs, categories (e.g. ``order \#12345'', ``electronics category'', ``user john\_doe'', ``product X123'')
- logic constraints: Boolean conditions, dependencies (e.g. ``only if'', ``unless'', ``when'', ``except'', ``provided that'')
- other condition constraints: Constraints that do not belong to the above categories
Rules:
1. Be literal: Extract exact words/phrases from objective - don't interpret or infer beyond what's stated
2. Categorize precisely: Each constraint MUST be assigned to exactly one category above
3. Include implicit: If objective implies a constraint (e.g., ``find the best'' implies ranking), include it
4. No duplicates: Same constraint should only appear once, even if mentioned multiple times
5. Do not infer the completion status of tasks; this section is only for information extraction.

~
INTERACTION HISTORY SUMMARY:
Provide a precise step-by-step summary of EVERY interaction shown in the INTERACTION HISTORY section. This summary MUST maintain perfect numerical alignment with the actual history steps.
Strict Constraints:
1. Full Sequential Summary: You must summarize every step in the history in chronological order. It is strictly forbidden to skip any steps, even if a step is repetitive, failed, or part of a loop.
2. Pure Objective Description: This section only allows factual statements. It is strictly prohibited to include any reasoning, subjective thoughts, or future plans. Simply describe what action was taken and what result was produced.
3. Action Summarization, Not Copying: Use concise language to summarize the intent of the action. It is strictly forbidden to directly copy the original action commands (e.g., click [123]); instead, describe it as ``Clicked the Add to Cart button.''
4. Historical Scope Only: Only summarize steps that have already been completed (Step 1 to Step N-1). Do not summarize the current step you are performing.
5. Retain Core Details: Ensure you capture key details from each step that affect the current state (e.g., specific values entered, specific filters selected, etc.).
~
CRITICAL STEP NUMBERING RULES:
- Count the EXACT number of <step\_X\_interaction> blocks in the INTERACTION HISTORY section
- If you see <step\_0\_interaction>, <step\_1\_interaction>, <step\_2\_interaction>, then summarize as ``Step 0:'', ``Step 1:'', ``Step 2:''
- If the history contains steps 0 through N-1, your summary MUST contain exactly steps 0 through N-1
- NEVER skip step numbers, NEVER start from Step 1 if Step 0 exists, NEVER add extra steps
Mandatory Verification Process:
1. Count History Steps: Identify all <step\_X\_interaction> blocks and note their exact numbers
2. Match Step Numbers: Your summary step numbers MUST exactly match the history step numbers
3. Complete Coverage: Summarize every single step shown in the history, no exceptions
Content Requirements:
- Factual Only: Pure objective description of what action was taken and what result occurred
- Concise Language: Summarize the intent, not the raw commands (e.g., ``Clicked the search button'' not ``click [123]'')
- Key Details: Include critical information that affects current state (values entered, pages navigated to, etc.)
- No Future Plans: Only describe completed actions, never predict or suggest next steps
Exact output format:
If current step = 0:
No history interactions available. This is the beginning of the task.
If current step > 0:
Step X: [Concise factual summary of action and result]
Step Y: [Concise factual summary of action and result]
...
(Where X, Y match the exact step numbers from INTERACTION HISTORY)

~
OBSERVATION DESCRIPTION:
Describe information in the CURRENT OBSERVATION section. Emphasize elements and features that are relevant or potentially helpful for fulfilling the objective in detail.
Strict Constraints:
1. Zero Reasoning: It is strictly prohibited to include any form of reasoning, interpretation, or assessment of task completion. Any inference in this section, however small, will negatively bias subsequent reasoning steps.
2. Precise Element Referencing: You MUST mention the specific element names along with their corresponding BIDs (e.g., [123]). This is critical for accurate element localization.
3. Pure Factual Description: Focus solely on describing the visible layout, text content, and interactive elements exactly as they appear on the page.

~
MODE ANALYSIS:
Analyze the current situation in MODE\_ANALYSIS section to determine the operational mode. Base your analysis STRICTLY on the information provided in the previous context and observations. Your analysis MUST be a step-by-step reasoning process that answers the following questions before making a final conclusion:

1. Current Page vs. Objective: Is the current page directly useful for achieving the task objective? Does it contain the necessary tools (e.g., date filters, specific input fields, detailed data tables) required by the constraints?

2. Task Progress: Where are we in the overall task? Is this the very first step? Or have we just completed a sub-task and are about to start a new one?

3. Information Sufficiency: Based on the provided context, do we have enough information about the current page to proceed with specific actions, or do we need to explore and gather more information first?
~
CRITICAL CONSTRAINTS:

- You MUST base your mode analysis ONLY on the information already provided in the context

- You are FORBIDDEN from reasoning about or suggesting specific actions to execute

- Your role is SOLELY to predict which operational mode (EXPLORATION or LOW-LEVEL ACTION) would be most appropriate for the next phase of task completion

- Focus on mode prediction, NOT action planning

After your step-by-step reasoning, you MUST conclude your analysis on a new line with the format: Conclusion: The current mode is [MODE], where [MODE] is either EXPLORATION or LOW-LEVEL ACTION.

~
REASON:
Provide your rationale for proposing the subsequent action commands here.
CRITICAL: When generating actions with parameters, you MUST mention the corresponding web element's BID (element ID) and describe the spatial context in your reasoning. This ensures accurate action execution.
Key points:
- Always mention the BID (e.g., [3164]) in your reasoning
- Describe the spatial context (e.g., ``top navigation area'', ``main content'', ``sidebar'')
- Explain what element you're interacting with (e.g., ``Search GitLab textbox'')
e.g. Objective is ``most recent'', but UI is ``Sort by: Hot''. Reasoning: ``Current view is sorted by popularity, not time. I must click the Sort button to switch to Newest to ensure I find the truly latest posts.''

~
ACTION:
Select your action here.

~
\end{promptbox}

\subsection{Action Space Definition}
\label{sec:action_definition}
This section provides detailed action definitions and associated prompts for WebNavigator introduced in \cref{sec:method_agent_design}.
Our agent operates with six actions in total. The high-level \texttt{navigate(domain, query)} action discussed in the methodology is implemented \texttt{optimal\_navigate[think][domain][query]}. \cref{table:action_space_comparison} compares our action space with prior methods.

\begin{promptbox}[Action Space Definition]{specifications}
CLICK:
click [id]: To click on an element with its numerical ID on the webpage. E.g., click [7] If clicking on a specific element doesn't trigger the transition to your desired web state, this is due to the element's lack of interactivity or GUI visibility. In such cases, move on to interact with OTHER similar or relevant elements INSTEAD.

~
TYPE:
type [id] [content] [press\_enter\_after=0|1]: To type content into a field with a specific ID. By default, the ``Enter'' key is pressed after typing unless press\_enter\_after is set to 0. E.g., type [15] [Carnegie Mellon University] [1]. If you can't find what you're looking for on your first attempt, consider refining your search keywords by breaking them down or trying related terms. To search for a GitLab user: type [2556] [user] [1]. To search for a project: type [2556] [project name] [1]. For example, type [2556] [keycloak/keycloak] [1]. Combining a search for an author and a project will fail. Identify what you're searching for (user, project, issue, etc.) and search for only that type of information.

~
STOP:
stop [answer]: To stop interaction and return response. Present your answer within the brackets. If the task doesn't require a textual answer or appears insurmountable, indicate ``N/A'' and additional reasons and all relevant information you gather as the answer. E.g., stop [5h 47min]
STRICT RULE: Use stop exclusively for final answers. Never use it for intermediate notes; use the note action for that purpose.

~
NOTE:
note [content]: To take note of all important info w.r.t. completing the task to enable reviewing it later. E.g., note [Spent \$10 on 4/1/2024]

~
GO BACK:
go\_back: To return to the previously viewed page.

~
OPTIMAL NAVIGATE
optimal\_navigate [thinking][domain][query]:
\# Action definition:
Teleports the agent to a specific webpage by performing a visual-semantic search over pre-indexed screenshots of the website. Use this to jump between different functional areas (e.g., from Dashboard to Settings) without repeatedly clicking.

\# Argument Construction:
[think]:
You must TRANSLATE the user's task into a VISUAL PAGE DESCRIPTION. Since the retrieval tool works by matching screenshots, you must simulate the visual environment of the target page.
\#\# Thinking Requirements (CRITICAL):
1.  **Length \& Detail**: The thinking block MUST be at least 50 words. Do not be concise.
2.  **Context Integration**: You must explicitly analyze the **User's Objective** (Key entities, IDs, dates) and the **Current Observation** (Why is the current page not good enough?).
3.  **Visual Simulation**: Before generating the query, vividly describe what you expect to see on the target page. Mention specific UI elements like ``tables with Status column'', ``forms with Refund button'', ``Sidebars with Analytics tab''.

\#\# Examples:
1. User Task: ``Check Reddit for news then post on Twitter.'' Current Goal: Check Reddit. 
  Output: optimal\_navigate [The immediate sub-goal is to browse news on Reddit. I am currently not on the reddit domain. Based on the user's request to Check Reddit, the correct domain is reddit. I need to jump to the specific subreddit listing page. The target page is not a specific post, but a feed. I expect to see the standard Reddit layout: a header with r/news, a feed of post titles, upvote arrows, and thumbnails.] [reddit] [r\/news subreddit listing page with top stories]
  
[domain]:
Select the domain parameter based on the website required for the current task. For tasks that span multiple websites, you will need to use different domains for different sub-tasks, which means you must break down the overall mission appropriately.

[query]: 
The visual-semantic search string describing the TARGET PAGE STATE.
- Good: ``Order \#555 details page with refund option'' (Describes the place)
- Bad: ``Refund order \#555'' (Describes the action)
The query must include the critical conditions and entities from the source objection. If an entity needs to be inferred, first deduce a more suitable entity and its corresponding conditions in the [thinking] section before including them in the query. This will significantly enhance retrieval effectiveness.

\# Output format: 
- You must strictly follow this format with NO extra text:
optimal\_navigate [Your detailed visual reasoning (min 50 words) analyzing objective vs observation] [domain] [The visual-semantic description of the page]

\end{promptbox}

\subsection{Selector prompt}
\label{sec:selector_prompt}
As the reasoning component within the Global-View Navigator, the Selector performs a three-step selection process on retrieved candidates: (1) Information sufficiency Analysis, (2) Relevance Filtering, and (3) Operational Efficiency Ranking. 


\begin{promptbox}[Selector Prompt]{selector}
You are a Goal-Oriented Navigation Module for an intelligent Web Agent. Your primary directive is to select a webpage that provides the most direct path to achieving the user's ultimate goal, prioritizing data completeness over superficial UI convenience.

Information: You will receive three types of information:
user\_objective: The user's final, high-level task goal. This is your strategic compass.
retrieval\_query: The description of intermediate steps generated by the Agent. This is your tactical instruction.
candidate\_pages: A series of webpage screenshots retrieved based on the retrieval\_query.
Core Task: Your decision process is a strict, hierarchical three-step evaluation:
Data Sufficiency Analysis (Highest Priority): First, evaluate each candidate page against the user\_objective. Determine if the data presented on the page is complete and sufficient to answer the user's final question. A page showing a subset of data (e.g., a Shipped Orders page when the goal is to find a customer's *entire order history*, or a Top 10 Customers report when the goal is to find a specific customer by name) is considered insufficient and must be deprioritized.

CRITICAL: If NONE of the candidate pages provide sufficient data or match the user's objective, you must strictly decide to return ``None''. Do not force a selection from irrelevant or incomplete pages.
Relevance Filtering: Among the pages deemed to have sufficient data, use the retrieval\_query as a benchmark to filter for pages that have the necessary UI elements and functionality for the next action.
Operational Efficiency Ranking: Finally, from the remaining candidates, choose the one that allows you to complete the user\_objective with the fewest subsequent operations. Remember, a path leading to an incomplete or incorrect answer is infinitely inefficient.

Reasoning Requirements:
At the beginning of your reasoning, you must explicitly restate the user\_objective and retrieval\_query.
Your analysis for each key candidate must follow the three-step evaluation (Sufficiency -> Relevance -> Efficiency). You must explicitly state your conclusion about the data sufficiency of each page.
 Clearly compare the pages, explaining why one is chosen over others based on this hierarchy. For instance:1) While page A is titled Products on Sale and has a very prominent search bar, its data is limited only to discounted items. This makes it insufficient for the user's objective of finding the price of *any* product. Page B, titled Product Catalog, contains all products and is therefore the correct choice, even if its search function is less obvious.
2) User Goal: Post in /f/relationship\_advice. Comparison: 1. Page A (titled Create new page) is rejected immediately because it is for creating Wikis, not user posts. 2. Page B (titled Create submission) is relevant but inefficient. Its Forum dropdown shows Choose one..., meaning the agent must manually search and select the subreddit (2 extra steps). 3. Page C (the /f/relationship\_advice forum index) is the optimal target. By selecting this page, the subsequent Submit action will inherit the current context and auto-fill the forum parameter, achieving the goal with minimum operations.
If no suitable page is found, explicitly explain why all candidates failed the Data Sufficiency or Relevance criteria.

Strict Output Format: Please output strictly according to the following JSON format.

\{ ``reasoning'': ``your reasoning content'', ``target\_page'': ``image name OR None''\}

JSON Format Mandatory Requirements - Must Be Strictly Followed:
The value of the reasoning field must be a continuous line of text, without any line breaks. Do not use Chinese quotation marks; if quoting is needed, use English single quotes.
Do not use backslash escape characters in strings. The target\_page value must be the specific image name if a suitable page is found, or the string ``None'' if no page meets the criteria.
Output pure JSON directly, without any other content, and do not wrap it in markdown code blocks.
\end{promptbox}


\subsection{LLM-based Exploration Prompt}
\label{sec:llm_explore_prompt}

\begin{promptbox}[LLM-based Web Exploration]{llm explore}
You are a Web Exploration Assistant. You are part of a system that constructs an Interaction Graph to capture the topological structure of a website. Your task is to determine the appropriate action to perform on a given web element in order to explore new page states and maximize the discovery of functional entry points.

\textbf{Core Principles}

Different types of elements require specific interaction methods to trigger changes in the page state:
\begin{itemize}
  \item Searchboxes / Input fields: Require a \texttt{type} action to enter content, followed by a \texttt{press} action (Enter) to trigger the search or submission.
  \item Comboboxes (Dropdowns): Can be selected to choose from provided child options, or typed into to filter options.
\end{itemize}

\textbf{Available Actions}
\begin{enumerate}
  \item \texttt{select} --- Select an option. params: \{"value": "option label"\}
  \item \texttt{type} --- Enter text into an input field. params: \{"value": "text to input"\}
  \item \texttt{press} --- Press a keyboard key. params: \{"key": "Enter"\} or \{"key": "Tab"\}
\end{enumerate}

\textbf{Action Generation Rules}

Determine the action sequence based on the element's role and attributes:

\texttt{searchbox}: Type a reasonable search term relevant to the website's content, then press Enter.

\texttt{textbox} (search-related): Type a search term, then press Enter. (standard input): Type \texttt{"test"}, then press Enter.

\texttt{combobox} (supports text filtering): Type a filter term, then press Enter. (pure dropdown): Select an available option label.

\textbf{Output Format}

Return the result in JSON format:
\begin{verbatim}
{
  "reason": "A brief explanation",
  "actions": [{"type": "action_type", "params": {parameters}}]
}
\end{verbatim}

\textbf{Examples}

Input:
\begin{verbatim}
{"role": "searchbox", "placeholder": "Search products...", "name": "Search"}
\end{verbatim}
Output:
\begin{verbatim}
{
  "reason": "Search boxes require typing followed by pressing Enter",
  "actions": [
    {"type": "type", "params": {"value": "shoes"}},
    {"type": "press", "params": {"key": "Enter"}}
  ]
}
\end{verbatim}

Input:
\begin{verbatim}
{"role": "combobox", "placeholder": "Select category",
 "name": "Category", "child_options": ["games", "foods"]}
\end{verbatim}
Output:
\begin{verbatim}
{
  "reason": "Select an available option",
  "actions": [
    {"type": "select", "params": {"value": "foods"}}
  ]
}
\end{verbatim}

\textbf{Input}

Page URL: \texttt{\{page\_url\}}

Element description: \texttt{\{element\_desc\}}
\end{promptbox}

\newif\ifincludechecklist
\includechecklistfalse
\ifincludechecklist
  \clearpage
  \section*{NeurIPS Paper Checklist}



\begin{enumerate}

\item {\bf Claims}
    \item[] Question: Do the main claims made in the abstract and introduction accurately reflect the paper's contributions and scope?
    \item[] Answer: \answerYes{} 
    \item[] Justification: Abstract and \cref{sec:intro}

\item {\bf Limitations}
    \item[] Question: Does the paper discuss the limitations of the work performed by the authors?
    \item[] Answer: \answerYes{} 
    \item[] Justification: \cref{sec:limitations}

\item {\bf Theory assumptions and proofs}
    \item[] Question: For each theoretical result, does the paper provide the full set of assumptions and a complete (and correct) proof?
    \item[] Answer: \answerNA{} 
    \item[] Justification: This paper does not include theoretical results.

    \item {\bf Experimental result reproducibility}
    \item[] Question: Does the paper fully disclose all the information needed to reproduce the main experimental results of the paper to the extent that it affects the main claims and/or conclusions of the paper (regardless of whether the code and data are provided or not)?
    \item[] Answer: \answerYes{}{} 
    \item[] Justification: The method is described in detail in \cref{sec:method}, with the full exploration algorithm provided in \cref{appendix:algorithm} and all hyperparameters and implementation details reported in \cref{sec:exp_setting}.

\item {\bf Open access to data and code}
    \item[] Question: Does the paper provide open access to the data and code, with sufficient instructions to faithfully reproduce the main experimental results, as described in supplemental material?
    \item[] Answer: \answerYes{} 
    \item[] Justification: The source code and pre-constructed Interaction Graphs are publicly available at \url{https://github.com/fate-ubw/webNavigator}, together with instructions for reproducing the main experimental results.

\item {\bf Experimental setting/details}
    \item[] Question: Does the paper specify all the training and test details (e.g., data splits, hyperparameters, how they were chosen, type of optimizer) necessary to understand the results?
    \item[] Answer: \answerYes{} 
    \item[] Justification: \cref{sec:exp_setting}

\item {\bf Experiment statistical significance}
    \item[] Question: Does the paper report error bars suitably and correctly defined or other appropriate information about the statistical significance of the experiments?
    \item[] Answer: \answerNo{} 
    \item[] Justification: Following standard practice in web agent evaluation, all experiments are conducted under the official evaluation protocols of WebArena\cite{webarena2023} and Online-Mind2Web\cite{onlinemind2web}, which are the established benchmarks in this field.

\item {\bf Experiments compute resources}
    \item[] Question: For each experiment, does the paper provide sufficient information on the computer resources (type of compute workers, memory, time of execution) needed to reproduce the experiments?
    \item[] Answer: \answerYes{}{} 
    \item[] Justification: \cref{sec:exp_setting}
    
\item {\bf Code of ethics}
    \item[] Question: Does the research conducted in the paper conform, in every respect, with the NeurIPS Code of Ethics \url{https://neurips.cc/public/EthicsGuidelines}?
    \item[] Answer: \answerYes{} 
    \item[] Justification: We have carefully read the code of ethics.

\item {\bf Broader impacts}
    \item[] Question: Does the paper discuss both potential positive societal impacts and negative societal impacts of the work performed?
    \item[] Answer: \answerYes{} 
    \item[] Justification: \cref{sec:impact}
    
\item {\bf Safeguards}
    \item[] Question: Does the paper describe safeguards that have been put in place for responsible release of data or models that have a high risk for misuse (e.g., pre-trained language models, image generators, or scraped datasets)?
    \item[] Answer: \answerYes{} 
    \item[] Justification:  \cref{sec:impact}

\item {\bf Licenses for existing assets}
    \item[] Question: Are the creators or original owners of assets (e.g., code, data, models), used in the paper, properly credited and are the license and terms of use explicitly mentioned and properly respected?
    \item[] Answer: \answerYes{} 
    \item[] Justification: All assets used in the paper are credited and the license and terms of use are properly respected.

\item {\bf New assets}
    \item[] Question: Are new assets introduced in the paper well documented and is the documentation provided alongside the assets?
    \item[] Answer: \answerYes{} 
    \item[] Justification: The released source code and pre-constructed Interaction Graphs are documented in the public repository, including setup and usage instructions.

\item {\bf Crowdsourcing and research with human subjects}
    \item[] Question: For crowdsourcing experiments and research with human subjects, does the paper include the full text of instructions given to participants and screenshots, if applicable, as well as details about compensation (if any)? 
    \item[] Answer: \answerNA{} 
    \item[] Justification: All manual annotations in this work were performed by the paper's authors; the study does not involve crowdsourcing or external human subjects.

\item {\bf Institutional review board (IRB) approvals or equivalent for research with human subjects}
    \item[] Question: Does the paper describe potential risks incurred by study participants, whether such risks were disclosed to the subjects, and whether Institutional Review Board (IRB) approvals (or an equivalent approval/review based on the requirements of your country or institution) were obtained?
    \item[] Answer: \answerNA{} 
    \item[] Justification: All manual annotations in this work were performed by the paper's authors; the study does not involve external human subjects.

\item {\bf Declaration of LLM usage}
    \item[] Question: Does the paper describe the usage of LLMs if it is an important, original, or non-standard component of the core methods in this research? Note that if the LLM is used only for writing, editing, or formatting purposes and does \emph{not} impact the core methodology, scientific rigor, or originality of the research, declaration is not required.
    \item[] Answer: \answerYes{} 
    \item[] Justification: LLMs are a core component of WebNavigator.

\end{enumerate}

\fi

\end{document}